\documentclass[12pt,a4paper]{article}

\usepackage[english]{babel}

\usepackage{yhmath}
\usepackage{amsfonts}
\usepackage{amsmath}
\usepackage[single]{accents}

\usepackage{array}
\usepackage{verbatim}
\usepackage{graphicx}
\usepackage{graphics}
\usepackage{subfigure}
\usepackage{epsfig}
\usepackage{tabularx}
\usepackage{mathcomp}
\usepackage{url}

\usepackage{booktabs}
\usepackage[T1]{fontenc}
\usepackage[utf8]{inputenc}

\newcommand{\mG}{\mathcal{G}}

\newcommand{\mS}{\mathcal{S}}
\newcommand{\mT}{\mathcal{T}}

\newcommand{\mV}{\mathcal{V}}

\newcommand{\Sij}{\mathcal{S}_{ij}}
\newcommand{\Skk}{\mathcal{S}_{kk}}

\newcommand{\bu}{\mathbf{u}}

\newcommand{\bF}{\mathbf{F}}

\newcommand{\bmG}{\boldsymbol{\mG}}

\newcommand{\bS}{\mathbf{S}}

\newcommand{\bU}{\mathbf{U}}

\newcommand{\bq}{\mathbf{q}}

\newcommand{\bbar}{\overline}

\newcommand{\frho}{\bbar{\rho}}

\newcommand{\fp}{\bbar{p}}

\newcommand{\wt}{\widetilde}

\newcommand{\fu}{\wt{u}}
\newcommand{\fh}{\wt{h}}
\newcommand{\fe}{\wt{e}}

\newcommand{\fT}{\wt{T}}

\newcommand{\fS}{\wt{\mathcal{S}}}
\newcommand{\fSij}{\wt{\mS}_{ij}}

\newcommand{\fSkk}{\wt{\mS}_{kk}}

\newcommand{\fsigmaij}{\wt{\sigma}_{ij}}

\newcommand{\de}{\partial}
\newcommand{\Div}{\nabla\cdot}
\newcommand{\Grad}{\nabla}

\newcommand{\sigmaij}{\sigma_{ij}}
\newcommand{\taukk}{\tau_{kk}}
\newcommand{\tauij}{\tau_{ij}}
\newcommand{\ei}{e_{\rm i}}

\newcommand{\bn}{\mathbf{n}}
\newcommand{\bx}{\mathbf{x}}
\newcommand{\br}{\mathbf{r}}

\newcommand{\bdelta}{\bbar{\Delta}}

\newcommand{\deltaij}{\delta_{ij}}
\newcommand{\nusgs}{\nu^{{\rm sgs}}}

\newcommand{\RE}{Re}
\newcommand{\MA}{M\hspace{-1pt}a}
\newcommand{\PR}{Pr}
\newcommand{\dm}{{\rm d}}

\newcommand{\rvect}{\pmb{r}}
\newcommand{\xvect}{\pmb{x}}

\newcommand{\mean}[1]{\left\langle #1 \right\rangle}
\newcommand{\tr}[1]{\operatorname{tr}\left(#1\right)}
\newcommand\norm[1]{\left\lVert#1\right\rVert}

\begin{document}

\title{A locally $p-$adaptive approach for Large Eddy Simulation of 
compressible flows in a DG framework}

\author{Matteo Tugnoli$^{(1)}$, Antonella Abb\`a $^{(1)}$, \\
Luca Bonaventura$^ {(2)}$, Marco Restelli$^{(3)}$}

\maketitle

 \begin{center}
{\small
(1) Dipartimento di Ingegneria Aerospaziale, Politecnico di Milano \\
Via La Masa 34, 20156 Milano, Italy\\
{\tt antonella.abba@polimi.it, matteo.tugnoli@polimi.it}\\
{$ \ \ $ }\\
(2) MOX -- Modelling and Scientific Computing, \\
Dipartimento di Matematica, Politecnico di Milano \\
Via Bonardi 9, 20133 Milano, Italy\\
{\tt luca.bonaventura@polimi.it}\\
{$ \ \ $ }\\
(3) NMPP -- Numerische Methoden in der Plasmaphysik \\
Max--Planck--Institut f\"ur Plasmaphysik \\
Boltzmannstra\ss e 2, D-85748 Garching, Germany \\
{\tt  marco.restelli@ipp.mpg.de}\\
}
\end{center}

\date{}


\vspace*{0.5cm}


\vspace*{0.5cm}

\pagebreak

\abstract{We investigate the possibility of reducing the computational burden 
          of LES models by employing local polynomial degree adaptivity in the 
          framework of a high order DG method. A novel degree adaptation
           technique especially featured to be effective for LES applications 
           is proposed and its effectiveness is compared to that
          of other criteria already employed in the literature. The resulting
           locally adaptive approach allows to achieve significant
          reductions in computational cost of representative LES
           computations.}


\pagebreak

\section{Introduction}
\label{sec:intro} \indent

Discontinuous Galerkin (DG) methods have become increasingly popular for 
Large Eddy Simulation (LES)  and Direct Numerical Simulation (DNS) of 
turbulent flows over the last two decades, as shown by the large number of 
paper devoted to this topic, among which we mention for example 
 \cite{abba:2015}, \cite{beck:2014}, \cite{collis:2002b}, 
\cite{collis:2002a},  \cite{vanderbos:2007},  \cite{landmann:2008}, 
\cite{sengupta:2007}, \cite{uranga:2011}, \cite{vanderbos:2010}, 
\cite{wei:2011}. 
The remarkable accuracy of DG methods implies however that a larger number of 
degrees of freedom is employed with respect to continuous finite elements
approache, and, as a consequence, a higher computational cost is required.
A natural way to reduce such cost in a DG framework is adapting the spatial resolution.
In this paper, we investigate the possibility of reducing the computational 
burden of LES by employing local polynomial degree adaptivity, known as 
\textit{p}-adaptivity, in the framework 
of the high order DG-LES model presented in \cite{abba:2015}.
 
\textit{p}-adaptivity consists in employing a polynomial base of different
degrees in  different elements and the refinement or coarsening of the resolution relies on the increase or reduction of the base degree.
While this kind of adaptivity gives somewhat less freedom than other types of 
mesh adaptivity, it is easy to implement in a DG framework and does not 
require remeshing nor any significant computational overhead.
Degree adaptivity has been proposed in the seminal papers  
\cite{zienkiewicz:1983}, \cite{zienkiewicz:1991} and various $p-$adaptive 
approaches have been proposed in the literature for a number of applications,
see e.g.  \cite{oden:1999}, \cite{burbeau:2005}, \cite{demkowicz:1991}, 
\cite{demkowicz:2002}, 
\cite{eskilsson:2011} \cite{flaherty:1995},  \cite{houston:2005}, 
\cite{remacle:2003}, \cite{tumolo:2015}, \cite{tumolo:2013}.

In a non adaptive LES, the mesh defines {\it a priori} the spatial resolution and therefore
 the size of the resolvable turbulent scales, 
often with rather little insight on the flow conditions that  are to be 
simulated, especially in complex geometries and in absence of reference high resolution
simulations. 
As a result,  the spatial resolution will be insufficient in some  regions 
of the computational domain and excessive in others.

Where the resolution is insufficient, the main turbulent scales will not be 
adequately simulated, while in over-resolved regions an excessive computational 
effort is spent in approximating flows with few or no turbulent structures. 
Under such circumstances,   a locally degree adaptive  LES  should have the double 
aim of reducing the computational cost by removing excessive degrees of 
freedom in less turbulent areas, while maintaining  or increasing accuracy by 
keeping or adding degrees of freedom in more turbulent areas.

However, very specific issues arise
when applying degree adaptation criteria to LES.
Using adaptivity in other contexts where a convergence to an exact solution is 
possible and advisable, the criteria used to decide whether to refine or 
coarsen are suitable local error indicators. These  may range from  smoothness indicators
\cite{burbeau:2005} to relative weight of the solution expansion modes 
\cite{tumolo:2013} and
solution of complex  dual problems \cite{hoffman:2005}, \cite{hoffman:2006}.
Even though the  convergence of a LES solution can be achieved, unlike 
for RANS solutions, increasing the resolution in LES would ultimately 
lead to a DNS solution,  as pointed 
out by \cite{mitran:2001} and \cite[p.~378]{sagaut:2006}.
 Therefore,  it is not a reasonable aim, since it will simply remove 
the advantages of performing a simulation that is not completely resolved. 
 Instead, the final aim of adaptive LES, as stated e.g. in \cite{pope:2004}, 
should be to obtain a procedure in which the resolution of the discretization, 
and consequently the size of the LES filter, is  adapted so as to resolve only 
a prescribed amount of the turbulent scales,  while modelling the others. 

In addition to the previous considerations, it has to be also noted that changing the 
resolution of the discrete  problem changes the  LES filter scale.
For this reason the adaptation in a LES context should not be driven by   
error minimization of the error, 
but rather aim at the identification of the local resolution and filter scale 
distribution most suitable to effectively simulate a turbulent flow and its local features.

It is generally very difficult to adapt the local resolution to reproduce only 
a prescribed amount of the energy at the turbulent scales, for example by evaluating the resolved 
turbulent energy with respect to the unresolved one, as proposed in 
\cite{pope:2004}. This is mainly because the only way to have any insight on the   
unresolved scales is through the resolved ones. For this reason, it is still 
unclear how to drive adaptivity by prescribing that a certain fraction of the 
turbulent scales has to be resolved.
In the present work  we try instead to achieve adaptation in LES context by using
indicators that, rather than  estimating the local error, try to measure  if the local flow conditions are  
sufficiently well resolved for the LES model to effectively account for 
unresolved motions or  not.


With this aim, we propose a novel local degree adaptivity criterion, based 
on an approximation of the classical structure function, see e.g. 
\cite{pope:2000}. The structure function is an estimate of the
correlation between velocity values at different locations.
If the structure function is computed for  each element over distances of the 
order of the element size, large values
will indicate strong fluctuations inside that  element, highlighting
the need for more resolution to effectively simulate the flow conditions. 
Oppositely, smaller values will denote either negligible fluctuations, 
as in laminar conditions, or a very well resolved turbulent region, thus
highlighting instead an opportunity to reduce the resolution in that element. 
In addition, to account for conditions in which the subgrid model simulates 
well the turbulent conditions, the form that the structure function assumes in 
isotropic homogeneous turbulence has been subtracted from the indicator. 
This novel approach has been compared to a  more standard criterion,
based on the relative weight of the solution expansion modes, described e.g. 
in \cite{remacle:2003}, \cite{tumolo:2013}.
The two refinement criteria have been tested in  a statical adaptivity 
framework on the test case of the flow around a square section cylinder at 
$Re = 22000$ and $Ma = 0.3$, in which adaptation is performed based
on a preprocessing of previous run data. 

The results of $p-$adaptive simulations carried out with different adaptation criteria
have been compared to those obtained in reference simulations with constant
polynomial degree. Both indicators were capable of highlighting domain areas of major turbulent activity, but the indicator based on the evaluation of of the structure function proved itself more capable to lead to accurate results than the one based on the relative weight of the modal solution. However,
results obtained with the  novel indicator led to accurate results, 
comparable to those obtained with constant maximum polynomial degree, while reducing
the required computational effort  by approximately 60\%, which was not the case for 
adaptation criteria based on error estimation.

 Furthermore, the sensitivity of the results to the resolution
of the preliminary run required to compute the adaptation indicator in a statically
$p-$adaptive approach has been studied. It was shown that
an accurate adaptive solution can be achieved using a relatively coarse resolution in the preliminary runs, thus outlining a practical procedure to obtain efficient adaptive results with minimal additional effort.

In section \ref{sec:dgles}, the compressible LES model based on a  DG 
discretization proposed in \cite{abba:2015} is reviewed. In section 
\ref{sec:adapt}, the adaptation criteria  are described. 
In section \ref{sec:results} the results of our numerical experiments are 
reported.
Finally, some conclusions and perspectives for future work are presented in section 
\ref{sec:conclu}.

\section{The compressible LES  DG model}
\label{sec:dgles} \indent


To model a turbulent compressible flow, we consider the compressible 
Navier--Stokes equations, already in non dimensional form
\begin{subequations}
\label{eq:nscompr}
\begin{align}
&\de_t \rho + \de_j (\rho u_j) = 0 \\
&\de_t (\rho u_i) + \de_j (\rho u_i u_j) + 
\frac{1}{\gamma\,\MA^2}\de_i p - \frac{1}{\RE}\de_j \sigmaij =
\rho f_i \label{eq:nscompr-momentum} \\
&\de_t (\rho e) + \de_j (\rho h u_j)
- \frac{\gamma\,\MA^2}{\RE} \de_j (u_i \sigmaij) \nonumber \\
&\hspace{3cm}+ \frac{1}{\kappa\RE\PR}\de_j q_j = \gamma\MA^2\rho f_j u_j
\label{eq:nscompr-energy},
\end{align}
\end{subequations}
where the variables are density $\rho$, momentum $\rho\bu$ and volume specific 
total energy $\rho e$. 
The non dimensional form of the Navier--Stokes is obtained by assuming a 
reference length $L_r$, density $\rho_r$,  velocity $V_r$ and temperature 
$T_r$, from which all other reference value are calculated.  
The Mach and Reynolds number are defined as
\begin{equation}
\MA = \frac{V_r}{\left( \gamma R T_r \right)^{1/2}}, \qquad
\RE = \frac{\rho_rV_rL_r}{\mu_r},
\label{eq:adim-numbers}
\end{equation}
while  other non dimensional numbers based on the gas specific heat and 
ideal gas constant are
\begin{equation}
\gamma = \frac{c_p}{c_v}, \qquad
\kappa = \frac{R}{c_p}.
\end{equation}
In equations \eqref{eq:nscompr} $p$ denotes the pressure, $\mathbf{f}$ a 
prescribed forcing, 
$\rho h = \rho e + p$ the enthalpy and $\sigma$ and $\bq$ the momentum 
and heat diffusive fluxes, respectively. 
Equations \eqref{eq:nscompr} must be complemented by a (dimensionless) 
state equation
\begin{equation}
p = \rho T,
\label{eq:state-eq}
\end{equation}
where $T$ is the temperature, and the definition of specific total and 
internal energy based on temperature and velocity is
\begin{equation}
e = \ei + \frac{\gamma\MA^2}{2} u_ku_k,
\qquad \ei = \frac{1-\kappa}{\kappa} T.
\label{eq:ei}
\end{equation}
To complete the set of equations is enough to specify the constitutive 
equations:
\begin{equation}
\sigmaij = \mu \Sij^d, \qquad
q_i = -\mu\de_i T,
\label{eq:constitutive}
\end{equation}
where the rate of strain tensor is defined as
\begin{equation}
\Sij = \de_j u_i + \de_i u_j \qquad
\Sij^d = \Sij - \dfrac{1}{3}\Skk\delta_{ij}
\end{equation}
and the dynamic viscosity, according to Sutherland hypothesis, is
\begin{equation}
\mu(T) = T^\alpha.
\end{equation}


In order to obtain equations that describe the evolution only of the larger turbulent scales,
the Navier--Stokes equations \eqref{eq:nscompr} must be filtered with an 
appropriate filter denoted by  $\bbar{\cdot}$ characterized by a spatial scale 
$\bdelta$.
In the approach outlined in \cite{abba:2015}, that will be described shortly, 
the actual implementation of the filter is strongly related to the spatial 
discretization.
As customary in both compressible LES and RANS, the Favre filter operator 
$\wt{\cdot}$ is introduced, which is defined implicitly by the Favre 
decomposition and is
applied to the non linear terms composed by the density to avoid subgrid terms
in the continuity equation:
\begin{equation}\label{eqn:favre_decomp}
 \bbar{\rho u_i} = \frho \wt{u}_i, \qquad
 \bbar{\rho e} = \frho \wt{e}, \qquad 
 \bbar{\rho \ei} = \frho \wt{\ei}, \qquad
 \bbar{\rho h} = \frho \wt{h} = \frho \wt{e} + \bbar{p}.
\end{equation}
In the same way, the Favre decomposition for the temperature, taking into
account \eqref{eq:state-eq}, yields:
\begin{equation}
 \bbar{\rho T} = \frho \wt{T} = \bbar{p}.
\label{eq:state-eq-Favre}
\end{equation}
Finally, it is possible to define a Favre filtered version of
\eqref{eq:constitutive} 
\begin{equation}
\fsigmaij = \mu(\wt{T}) \fSij^d, \qquad
\wt{q}_i = -\mu(\wt{T}) \de_i \fT,
\label{eq:constitutive-Favre}
\end{equation} 
by neglecting small scale contributions, where $\fSij = \de_j \fu_i + \de_i 
\fu_j$ and $\fSij^d = \fSij -
\dfrac{1}{3}\fSkk\delta_{ij}$. 
After applying the filter to equations \eqref{eq:nscompr}, 
a number of subgrid terms arise from filtering the nonlinear 
terms and need to be modelled. Following 
\cite{abba:2015} a number of such subgrid terms are considered negligible and 
the resulting filtered equations containing only the relevant subgrid terms 
are:
\begin{subequations}
\label{eq:filteq}
\begin{align}
&\de_t \frho + \de_j (\frho \fu_j) = 0 \\
&\de_t \left( \frho \fu_i \right) + \de_j \left(\frho \fu_i \fu_j\right) 
+ \frac{1}{\gamma\,\MA^2}\de_i \fp -\frac{1}{\RE} \de_j \fsigmaij \nonumber \\
& \qquad \qquad = - \de_j \tauij + \frho f_i \label{eq:filteq-momentum} \\
& \de_t \left(\frho\fe\right) + \de_j \left(\frho\fh \fu_j\right) 
- \frac{\gamma\,\MA^2}{\RE}\de_j \left(\fu_i \fsigmaij \right)
+ \frac{1}{\kappa\RE\PR}\de_j \wt{q}_j  \nonumber  \\
& \qquad \qquad =
- \frac{1}{\kappa}\de_j Q_j^{{\rm sgs}}
- \frac{\gamma\MA^2}{2}\de_j \left( J_j^{{\rm sgs}} - \taukk\fu_j \right) 
\label{eq:filteq-energy} \\
&\quad\quad\quad + \gamma\MA^2\frho f_j \fu_j.   \nonumber 
\end{align}
\end{subequations}
where the contributions to be modelled are the subgrid stress tensor 
$\tauij = \bbar{\rho u_i u_j} - \frho\fu_i\fu_j$, the subgrid heat flux 
$Q_i^{{\rm sgs}}  = \bbar{\rho u_i T} - \frho\fu_i\fT$ and the turbulent 
diffusion flux $J_i^{{\rm sgs}}  = \bbar{\rho u_iu_ku_k} - 
\frho \fu_i\fu_k\fu_k .$


Concerning the subgrid stress model, we will focus on this work exclusively on 
the Smagorinsky model \cite{smagorinsky:1963}. Exploring the interaction 
between local degree adaptivity and more sophisticated subgrid scale models 
will be the target of future investigations.
The modelled terms are assumed to be proportional to 
the gradients of the variables,  with fixed
proportionality constants.  
In the Smagorinsky model, the deviatoric part of the subgrid stress tensor 
$\tau_{ij}$ in (\ref{eq:filteq}) is modelled by a so called eddy viscosity
$\nu^{{\rm sgs}}$, yielding
\begin{subequations}\label{eqn:nu_smag}
\begin{align}
 &  \tauij -\frac{1}{3}\tau_{kk} \deltaij = - \frac{1}{\RE} \frho 
 \nu^{{\rm sgs}} \fSij^d,
\label{eqn:nu_smag:tauij}
 \\
& \nu^{{\rm sgs}} = \RE\, C_S^2\bdelta^2 |\fS| f_D,
\label{eqn:nu_smag:nu}
\end{align}
\end{subequations}
where $C_S=0.1$ is the Smagorinsky constant, $|\fS|^2 =
\dfrac{1}{2}\fSij\fSij$ and $\bdelta$ is the filter length scale.
To account for the smaller size of the turbulent structures in the vicinity of 
a wall is necessary to correct the filter size $\bdelta$ using the function 
$f_D$, known as Van Driest damping function, in order to recover the correct 
trend of the turbulent viscosity, see for example \cite{sagaut:2006}. The 
function is defined as 
\begin{equation}
f_D(y^+) = 1 - \exp\left(- y^+/A \right),  
\end{equation}
where $A$ is a constant and $y^+=\frac{\rho_r u_\tau^\dm d^\dm_{{\rm
wall}}}{\mu_r}$, with $d^\dm_{{\rm wall}}$ denoting the (dimensional)
distance from the wall and $u_\tau^\dm$ the (dimensional) friction
velocity. In this work, a  value of $A=25$ has been considered.

Concerning the isotropic part of the subgrid stress tensor,
in the present work we have neglected it following other authors like 
\cite{erlebacher:1992} considering it negligible with respect to the pressure 
contribution.
The temperature flux $Q_i^{{\rm sgs}}$ from \eqref{eq:filteq-energy} is 
modelled with a  similar eddy viscosity concept, following \cite{eidson:1985}, 
as 
\begin{equation}\label{eqn:Qj_smag}
 Q_i^{{\rm sgs}} = - \frac{\PR}{\PR^{{\rm sgs}}} \frho \nusgs \de_i\fT,
\end{equation}
where $\PR^{{\rm sgs}}$ is a subgrid Prandtl number.
Finally, the turbulent diffusion flux $J_i^{{\rm sgs}}$ in 
(\ref{eq:filteq-energy}) can be rewritten 
using generalized central moments as in 
\cite{germano:1992}, and neglecting  the third order contribution in the 
velocity in analogy with RANS models,(see e.g. \cite{knight:1998}), yields
\begin{equation}\label{eqn:Jj_smag}
 J_i^{{\rm sgs}} \approx 2\fu_k\tau_{ik}.
\end{equation}

%
Equations \eqref{eq:filteq}, equipped with the appropriate subgrid stress 
model, are discretized in space by a discontinuous finite element approach, 
based on the so called Local Discontinuous Galerkin (LDG) method, see 
\cite{cockburn:1998b}, to approximate the 
second order derivative of the viscous terms.
The equations \eqref{eq:filteq} can be rewritten in compact form:
\begin{equation}
\label{eq:compact_form}
\de_t \bU + \Div \bF^{{\rm c}}(\bU) =  \Div \bF^{{\rm v}}(\bU,\Grad \bU) 
 -\Div \bF^{{\rm sgs}}(\bU,\Grad \bU) + \bS  
\end{equation}
where $\bU=[\frho\,,\frho\wt{\bu}^T,\frho\fe ]^T$ are the prognostic 
variables, $\bF^{{\rm c}}$ the convective fluxes, $\bF^{{\rm v}}$ and 
$\bF^{{\rm sgs}}$ the viscous and sub grid fluxes and $\bS$ the source term 
due to the generic forcing term.
Using the LDG method the gradients in \eqref{eq:compact_form} are substituted 
by an auxiliary variable $\bmG$ and an additional equation for the gradient
is introduced, obtaining 
\begin{eqnarray}
\label{eq:ldg_form}
\de_t \bU + \Div \bF^{{\rm c}}(\bU) &=&  \Div \bF^{{\rm v}}(\bU,\bmG) 
 -\Div \bF^{{\rm sgs}}(\bU,\bmG) + \bS   \\
 \bmG &-& \nabla{\boldsymbol \varphi} = 0,\nonumber
\end{eqnarray}
in which $\bmG$ represents the gradient of ${\boldsymbol \varphi} = 
[\wt{\bu}^T,\fT]^T$,
which are the only variables whose  gradients are used in the definition of 
viscous and turbulent fluxes.
The spatial discretization of \eqref{eq:ldg_form} starts from the definition
of a tessellation $\mT_h$ of the domain $\Omega$ into non overlapping 
tetrahedral elements $K$, over which a discontinuous polynomial finite element 
space $\mV_h$ is defined
\begin{equation}\label{eqn:mV_def}
\mV_h = \left\{ v_h \in L^2(\Omega): v_h|_K \in \mathbb{P}^{q_K}(K), \,
\forall K\in\mT_h \right\}.
\end{equation}
$\mathbb{P}^{q_K}(K)$ denotes the space of polynomial functions of total 
degree $q_K$, which can arbitrarily vary from element to element.
By defining the outward unit normal  $\bn_{\partial K}$ on the boundary of 
each element $\partial K$, and denoting with $(\bU_h,\bmG_h)\in(\,(\mV_h)^5\,,
\,(\mV_h)^{4\times3}\,)$ the numerical solution, it is possible to formulate
 the following weak problem from  equation \eqref{eq:ldg_form}
\begin{subequations}
\label{eq:DG-space-discretized}
\begin{align}
\displaystyle
\frac{d}{dt}\int_K \bU_h v_h\,d\bx
& \displaystyle
- \int_K \bF(\bU_h,\bmG_h)\cdot\nabla v_h\, d\bx
\\[3mm]
& \displaystyle
+ \int_{\partial K} \wideparen{\bF}(\bU_h,\bmG_h)\cdot \bn_{\partial K} v_h\,
d\sigma
= \int_K \bS v_h \,d\bx,\nonumber 
\\[3mm] \displaystyle
\int_K \bmG_h \cdot \br_h \,d\bx
& \displaystyle
+ \int_K {\boldsymbol \varphi_h}\nabla\cdot\br_h\, d\bx
\\[3mm]
& \displaystyle
- \int_{\partial K} \wideparen{{\boldsymbol \varphi}} \bn_{\partial
K}\cdot\br_h \, d\sigma = 0, \nonumber 
\\[3mm] \displaystyle
\forall K\in\mT_h, \quad &\forall v_h\in\mV_h, \quad \forall
\br_h\in(\mV_h)^3.\nonumber
\end{align}
\end{subequations}
where $\bU_h=\left[ \rho_h\,,\rho_h\bu_h\,,\rho_he_h \right]^T$ is the 
numerical solution in prognostic variables, 
${\boldsymbol \varphi}_h=\left[ \bu_h\,,T_h \right]^T$ is the numerical
counterpart of the gradient variables, $\bF =\bF^{{\rm c}}-
\bF^{{\rm v}}+\bF^{{\rm sgs}}$ denotes shortly the sum of all the fluxes and
$\wideparen{\bF}, $  $\wideparen{{\boldsymbol \varphi}}$ denote the  
numerical fluxes.
A numerical flux is needed since the solution  is not in general 
single valued at the element boundaries, due to the discontinuity of the 
finite element space.
This can be seen as a weak 
imposition of the solution on each element boundary, due to actual boundary 
conditions on external boundaries, or other elements solution on internal 
boundary between elements.
In the present work we employed a Rusanov flux for the convective flux 
$\wideparen{\bF}^{\rm c}$ and centred fluxes for viscous, subgrid fluxes 
$\wideparen{\bF}^{\rm v}$, $\wideparen{\bF}^{\rm sgs}$ and for gradient fluxes 
$\wideparen{{\boldsymbol\varphi}}$, whose detailed definitions can be found in \cite{giraldo:2008}.

As mentioned before, the definition of the filter operator $\bbar{\cdot}$  is 
based on the DG discretization, as proposed also in \cite{vanderbos:2007}. 
Let $\Pi_{\mV}:L^2(\Omega)\to\mV$  be the $L^2$ projector over 
the subspace  $\mV\subset L^2(\Omega)$, defined by  
\[
\int_\Omega \Pi_{\mV}u\,v\, d\bx =
\int_\Omega u\,v\, d\bx, \qquad \forall u,v \in\mV.
\]
The filter $\bbar{\cdot}$  is now simply defined as the projection over the 
solution subspace
\begin{equation}
\bbar{v} = \Pi_{\mV_h}v.
\label{eq:filter-bar}
\end{equation}
The application of this filter coincides with the projection of the solution 
over the finite dimensional numerical subspace $\mV_h.$ For this reason the 
filtered prognostic quantities $\frho$, $\frho\wt{\bu}$ and $\frho\fe$
can be identified with their numerical solution counterparts $\rho_h$, $\rho_h 
\bu_h $ and $\rho_he_h.$

The actual implementation of the discretization 
\eqref{eq:DG-space-discretized} used to compute the results presented in 
section \ref{sec:results} is based on \texttt{FEMilaro} \cite{femilaro}, a 
generic finite element library written using latest Fortran and MPI standards.
The hierarchical orthonormal basis obtained from the Legendre polynomials
has been employed for the 
subspace $\mV_h$, which has also been used to represent the prognostic 
unknowns, yielding a so called \emph{modal} DG formulation. 
As a result, for a generic model variable $\alpha, $ its
$ \mathbb{P}^{q_K}(K) $  numerical approximation
can be written as 
\begin{equation}
\label{modal_exp1}
\alpha_h|_K=\sum_{l=0}^{n_\phi(K)}\alpha^{(l)}\phi_{l}^{K},
\end{equation}
 where $\phi_{l}^{K}$ are the Legendre basis functions on element $K,$
 $\alpha^{(l)} $ the modal coefficients and $n_\phi(K)+1$ is the number
 of basis functions required to span $ \mathbb{P}^{q_K}(K).$
 Notice that, thanks to the hierarchical nature of the Legendre
 polynomial basis, equation \eqref{modal_exp1} can also be rearranged as
 \begin{equation}
 \label{modal_exp2}
 \alpha_h|_K=\sum_{p=0}^{q_K}\sum_{l\in d_p}\alpha^{(l)}\phi_{l}^{K},
\end{equation}
where $d_0=\left\lbrace 0\right\rbrace $ and
$$ d_p=\left\lbrace l\in 
1...n_\phi (K)\ \  | \ \  \phi_l\in \mathbb{P}^p(K) \backslash \mathbb{P}^{p-1}(K) \right\rbrace $$
is the set of indices of the basis functions of degree $p.$  
Notice that we choose the basis normalization in such a way that the coefficient
$\alpha^{(0)} $ coincides with the mean value of $ \alpha_h|_K $ over $K.$
 
All the integrals in 
\eqref{eq:DG-space-discretized} and in the projection \eqref{eq:filter-bar} have 
been computed using symmetrical quadrature rules from \cite{cools:2003} and 
\cite{zhang:2009}. The formulae employed must be exact at least up to the degree
$2q_K$ on each element to correctly integrate \eqref{eq:DG-space-discretized}. 
For this reason, the quadrature degree on the boundary of the element $\partial 
K$ should not be lower than $2q_K$ from both sides. To enforce this condition, the integrals 
on internal boundaries between two elements are performed with the maximum 
degree between the two elements. On the external 
boundaries, where a boundary condition is prescribed, the quadrature degree is 
simply taken as the internal one. 
Once discretized in space, the equations \eqref{eq:filteq} are advanced in time 
with an explicit,   five stages fourth order, Strong Stability Preserving 
Runge-Kutta method proposed in \cite{spiteri:2002}.

\section{The degree adaptation criteria}
\label{sec:adapt} \indent
 
The reasons for using $p-$adaptive techniques in the LES context
and  the  peculiarities of degree adaptation for such problems have been 
discussed in  section \ref{sec:intro}.
Here, the chosen adaptivity indicators   will be introduced. Notice that
degree adaptation can alternatively be performed either during the 
simulation (dynamic adaptivity), or only at the beginning of the simulation 
(static adaptivity).
In the present work, we  only present  results obtained in statically adaptive 
simulations. 
In this case, the results of a previous run are used to calculate an 
indicator, which is used to determine the polynomial degree distribution at 
the beginning of the simulation and then kept constant. 
This approach does not allow to adapt the resolution  in order to track 
non stationary phenomena, as done e.g. in  \cite{tumolo:2015}, 
\cite{tumolo:2013}, but it is nonetheless effective on statistically 
stationary flows and has the main advantage of allowing a simple load 
balancing of the simulation load on different processors without need of 
dynamic load balancing. 
We plan to investigate the dynamic adaptivity in future works. 

\subsection{Indicator based on modal coefficients}
\label{subs:RW indicator}
We consider first a simple indicator based on the comparison of a proxy of the
kinetic energy content at the smallest scales with respect to the one contained 
in all the scales.
The main advantage of this approach, that is analogous to the one employed in 
\cite{tumolo:2015}, \cite{tumolo:2013},   is that it only uses modal values, 
without the need for interpolation, thus being of very 
low computational impact. 
The indicator is defined, on each element, as 
\begin{equation}
\label{eq:ind_RW}
Ind_{M}(K) = \sqrt{\frac{e^s(K)}{e^{*}(K)}},
\end{equation}
where we define

$$ 
e^{*}(K) = \sum_{i=1}^3\int_K(\rho u_i)' (\rho u_i)'\ dv, \ \ \ 
 e^{s}(K) = \sum_{i=1}^3 \int_K(\rho u_i)^s (\rho u_i)^s\ dv.
$$
Here the superscript $^s$ indicates the smallest scale contribution and the 
prime $'$ indicates the value of momentum after  the mean value  over the 
element has been subtracted. 
Denoting by $m^{(l)}_i$ the modal coefficients of the $i-$th component
of momentum and using the representation in \eqref{modal_exp2},
the indicator
\eqref{eq:ind_RW} can be easily computed as 
\begin{equation}
e^{*}(K)= \sum_{i=1}^3 \sum_{l=1}^{n_\phi(K)} (m^{(l)}_i)^2,
\label{eq:rw_etot}
\end{equation}
\begin{equation}
e^{s}(K)=  \sum_{i=1}^3  \sum_{l\in d_{q_K}} (m^{(l)}_i)^2.
\end{equation}
The square of the momentum was employed in the indicator
instead of the kinetic energy because the modal expansion of
the momentum variables is immediately available and its use does not
entail computational overheads, while bringing similar information.
The mean over the element of the variable has been removed to avoid the 
presence of a mean field strongly variable in the domain to affect the 
indicator by under-weighting areas of strong mean flow, and to ensure Galilean invariance, as done in a similar context by \cite{flad:2016}. 

In a LES context, this indicator   can be interpreted as a rough local 
estimate of the relative amount of energy stored in smaller scales with 
respect to larger ones. 
Throughout the energy cascade and especially in the inertial range  where LES 
operates, the energy contained at smaller scales is less than that present  at 
the larger ones. 
A higher amount of energy at the small scales (and therefore in higher order 
modes) is an unphysical sign of insufficient resolution and ineffectiveness of 
the subgrid model.
Therefore, it should trigger a  resolution increase, while a very small amount 
of energy contained  at the finer 
scales simply means that they do not play a significant  role. 
While a similar indicator was employed in \cite{tumolo:2015}, \cite{tumolo:2013}
for dynamically adaptive simulations and computed at run time to adjust the
local degrees of freedom, for
static adaptivity the indicator value is computed from quantities
averaged in time over a preliminary simulation.

\subsection{Indicator based on the structure function}
\label{subs:SF indicator}
As an alternative to the indicator \eqref{eq:ind_RW}, we now consider 
a new indicator, meant to allow a better physical insight on the local flow 
conditions. 
The indicator is based on the classical structure function, which has been 
used extensively to study   turbulence statistics,
\begin{equation}
\label{eq:strucfun}
	D_{ij} = \mean{\left[u_i(\xvect+\rvect,t)-
	u_i(\xvect,t)\right]\left[u_j(\xvect+\rvect,t)-u_j(\xvect,t)\right]},
\end{equation}
where $\mean{\cdot}$ represents the expected value operator.
The structure function estimates the lack of correlation
in  the velocity values at the two points $\xvect+\rvect$ and $\xvect$. If 
evaluated inside each element, taking  the values of $\rvect$
 comparable   to the element size, it can estimate how much the solution 
is fluctuating inside each element. 
As previously remarked, large values of the structure function will  denote
the need for more resolution to effectively simulate the flow conditions, 
while smaller values will denote either   laminar conditions, or a very well 
resolved turbulent region, thus suggesting instead    the possibility of 
reducing the resolution in that element.  

It should be remarked, however, that  most of the subgrid models (and in 
particular the Smagorinsky model) perform adequately in a regime of 
homogeneous isotropic turbulence. 
Therefore, in such regimes excessive refinement is not necessary and one can 
let the subgrid scale  model simulate the turbulent dissipation. 
For this reason, we  remove from the structure function \eqref{eq:strucfun} 
the contribution due to homogeneous isotropic turbulence, for which the 
structure function becomes an isotropic function of the sole $\rvect$, i.e. 
$D_{ij}(B\rvect) = BD_{ij} (\rvect)B^T$ where $B$ is a rotation tensor. 
Using the notation of \cite{pope:2000} (p. 192), the structure function in isotropic 
conditions takes then the form
\begin{equation}
\label{eq:iso_strucfun}
	D_{ij}^{iso}(\rvect, t) = D_{NN}(r,t)\delta_{ij} +\left(D_{LL}(r,t) - D_{NN}
	(r,t)\right)\frac{r_ir_j}{r^2}
\end{equation}
where $r=\norm{\rvect}$  and $D_{LL}, D_{NN}$ are the
longitudinal and transverse structure functions, respectively.
Once $\rvect$ is  known, only  $D_{LL}$ and $D_{NN}$ 
 need to be determined. The complete calculations to obtain such values is presented in appendix \ref{app:anisotropy-calc}.
To obtain the isotropic form of the structure function and remove
it from the calculated structure function we use the following procedure:
\begin{enumerate}
\item choose a pair of points defining $\xvect$ and $\rvect$ in $K$
\item compute the structure function $D_{ij}(K)$ based on $\xvect$, $\rvect$ 
and the simulated velocity field
\item compute $D_{NN}$ and  $D_{LL}$ by a least square fit of   
\eqref{eq:iso_strucfun}
to the structure function values within the element
\item compute the square of the difference between the   structure function 
\eqref{eq:strucfun} and the approximate isotropic form 
\eqref{eq:iso_strucfun}:
\begin{equation}
\label{eq:qerr_def-app}
Q (K)=    \sum_{ij}\left[D_{ij}(K) - D_{ij}(K)^{iso}\right]^2.
\end{equation}
\end{enumerate}
Notice that the resulting scalar is the Frobenius norm of the structure 
function \eqref{eq:strucfun} minus the contribution that would be generated in 
case of homogeneous isotropic turbulence. The degree adaptation  indicator can 
then be defined as: 
\begin{equation}
\label{eq:ind_SF}
Ind_{SF}(K) = \sqrt{Q}(K) = \sqrt{\sum_{ij}\left[D_{ij}(K) - D_{ij}
(K)^{iso}\right]^2}.
\end{equation}

The structure function definition \eqref{eq:strucfun} relies on the 
expected value  of the correlation and an adequate approximation through some 
sort of averaging procedure of the correlation is required.  
For static adaptivity, the 
correlation is computed by averaging  over all the available previous results, 
while its anisotropic component is  evaluated by the previously described 
procedure.  
For dynamic adaptivity, a running time average of the structure function 
could be computed.
 
In order to reduce the arbitrariness in the choice of $\xvect$ and $\rvect$ for the
computation of  
\eqref{eq:strucfun}, its value is computed for each couple of element vertices 
and then averaged.
Alternatively, the maximum value among those obtained for each couple of 
element vertices could be considered. 
This procedure should create a sufficiently local quantification of the 
turbulence intensity, aimed at revealing where the turbulent flow features 
allow for a coarser resolution and more modelling.
The structure function indicator  \eqref{eq:ind_SF} could  give more physical 
insight on the flow conditions and be more robust than
the indicator \eqref{eq:ind_RW}, but it is more complex and implies a 
greater computational overhead, since, in our modal framework, it requires the reconstruction
 of the solution point values
at every element vertex.
Note that an indicator based on the interpolated values of the solution at different points of the element was already presented in \cite{burbeau:2005}, but in that case the values were used to evaluate a local approximation of the  velocity gradients.

\section{Numerical results}
\label{sec:results} \indent


To fully exploit the advantages of an adaptive LES framework on a simple 
geometry, we identified as test case the compressible flow around a square section cylinder 
at Mach number $Ma=0.3$ and
 Reynolds number  $Re = 22000$, based on free stream conditions and cylinder side. This test has been widely used 
as a benchmark for LES computations and for this reason a large amount of 
data, both numerical and experimental, are available for comparison. 
This flow is representative of  flows around bluff bodies with sharp 
edges and it is ideal to test adaptivity in LES, since  a variety of 
conditions are expected in the domain, from laminar far field to separations, big vortices 
and small scale turbulence. 
The flow separates immediately at the front edges of the 
square cylinder. On the upper and lower sides, recirculating bubbles are 
present, while on the rear of the cylinder a bigger recirculation zone is 
subject to the big vortices detached periodically from the upper and lower 
recirculation zones. The spanwise direction, along the length of the cylinder, 
can be considered statistically homogeneous. 


Several authors presented experimental results regarding this type of flow, 
some of which are grouped in table \ref{tab:exp-global}. While there is a 
strong agreement on the shedding frequency and its associated Strouhal number, 
as well as on the mean drag coefficient, there are less data and less agreement on the 
force coefficient fluctuations, which seem to depend heavily on the inflow 
turbulence level, see \cite{mclean:1992}. 
Regarding velocity statistics in proximity of the cylinder, the data from  
  \cite{lyn:1994}, \cite{lyn:1995} available from \cite{agard345} are used 
as a reference and will be cited as "experimental data" in the following figures.

\begin{table*}
\centering
\begin{tabular}{rccccc}
\toprule
            & $Re/10^3$      & St            & <Cd>     & rms(Cl')   \\ \midrule
Lyn et al. \cite{lyn:1994}, \cite{lyn:1995} & 21.4    & 0.13 & 2.1    & - \\
Norberg \cite{norberg:1993} & 1.3 and higher & 0.13 & 2.16   & - \\
Bearman and Obajasu \cite{bearman:1982} & 22 & 0.13 & 2.1  & 1.2 \\
McLean and Gartshore \cite{mclean:1992} & > 20 & -    & -   & 1.1-1.4 \\
Minguez et al. \cite{minguez:2011} & 22 & 0.13 & 2.1   & -
 \\

\bottomrule
\end{tabular}
\caption{Global results of different laboratory experiments}
\label{tab:exp-global}
\end{table*}

Several numerical simulations have also been performed on this test case by 
different authors, to test different LES approaches and numerical 
frameworks. The global results of some of 
them are presented in table \ref{tab:num-global}. It is evident that the 
results presented by those authors, while in most cases   plausible, 
are rather scattered
around the experimental values. This highlights how the test case is 
challenging and how difficult it is to identify a reference value to assess the 
quality of the simulations.

\begin{table*}
\centering
\begin{tabular}{rcccccc}
\toprule
            &  St            & <Cd> & rms(Cd')    & rms(Cl')   \\ \midrule
Koobus \& Farhat \cite{koobus:2004}          & 0.136 & 2.00  & 0.19  & 1.01 \\
Minguez et al. \cite{minguez:2011}           & 0.141 & 2.2   & -     & - \\
Minguez et al. \cite{minguez:2008}           & 0.141 & 2.31  & 0.131 & - \\
Murakami et al. \cite{murakami:1999}         & 0.135 & 1.99  & -     & 0.86 \\
Murakami \& Mochida \cite{murakami:1995}     & 0.132 & 2.09  & 0.13  & 1.60 \\
Sohankar \& Davidson \cite{sohankar:2000}    & 0.127 & 2.22  & 0.16  & 1.50 \\
Verstappen \& Veldman \cite{verstappen:1998} & 0.133 & 2.09  & 0.178 & 1.45 \\
Voke \cite{voke:1997}     & 0.13-0.161 & 2.041-2.79 & 0.12-0.36 & 1.01-1.68 \\
Rodi \cite{rodi:1997b}    & 0.066-0.15 & 1.66-2.77 & 0.10-0.27 & 0.38-1.79\\
Trias et al. \cite{trias:2015}               & 0.132 & 2.18 & 0.205  & 1.71 \\
\bottomrule
\end{tabular}
\caption{Global results of different numerical simulation at the same Reynolds 
number, $Re = 22000$ }
\label{tab:num-global}
\end{table*}

\begin{figure*}
\centering
\includegraphics[width=0.8\textwidth]{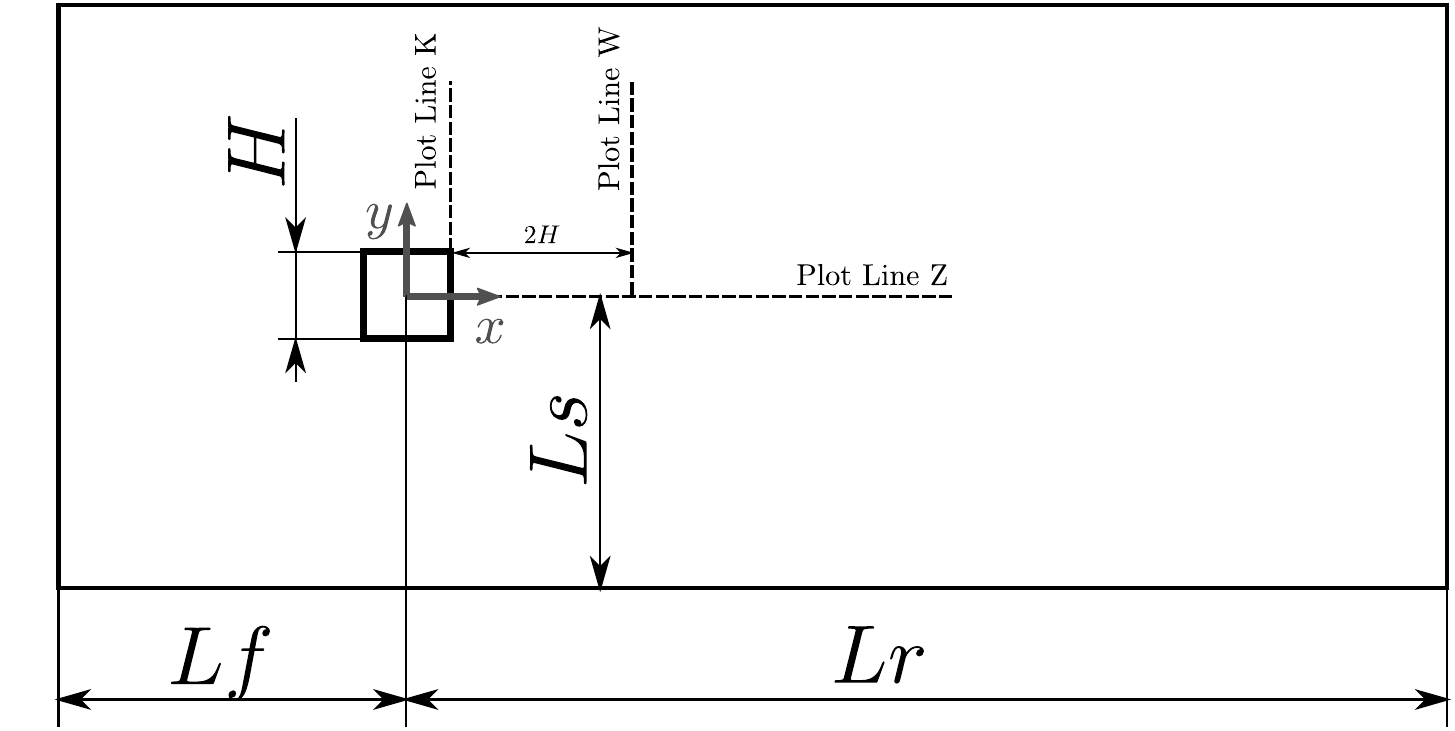}    
\caption{Sketch of the square section cylinder computational domain, side view, with plot lines}
\label{fig:cyl}
\end{figure*}

\begin{figure*}
\centering
\includegraphics[width=0.8\textwidth]{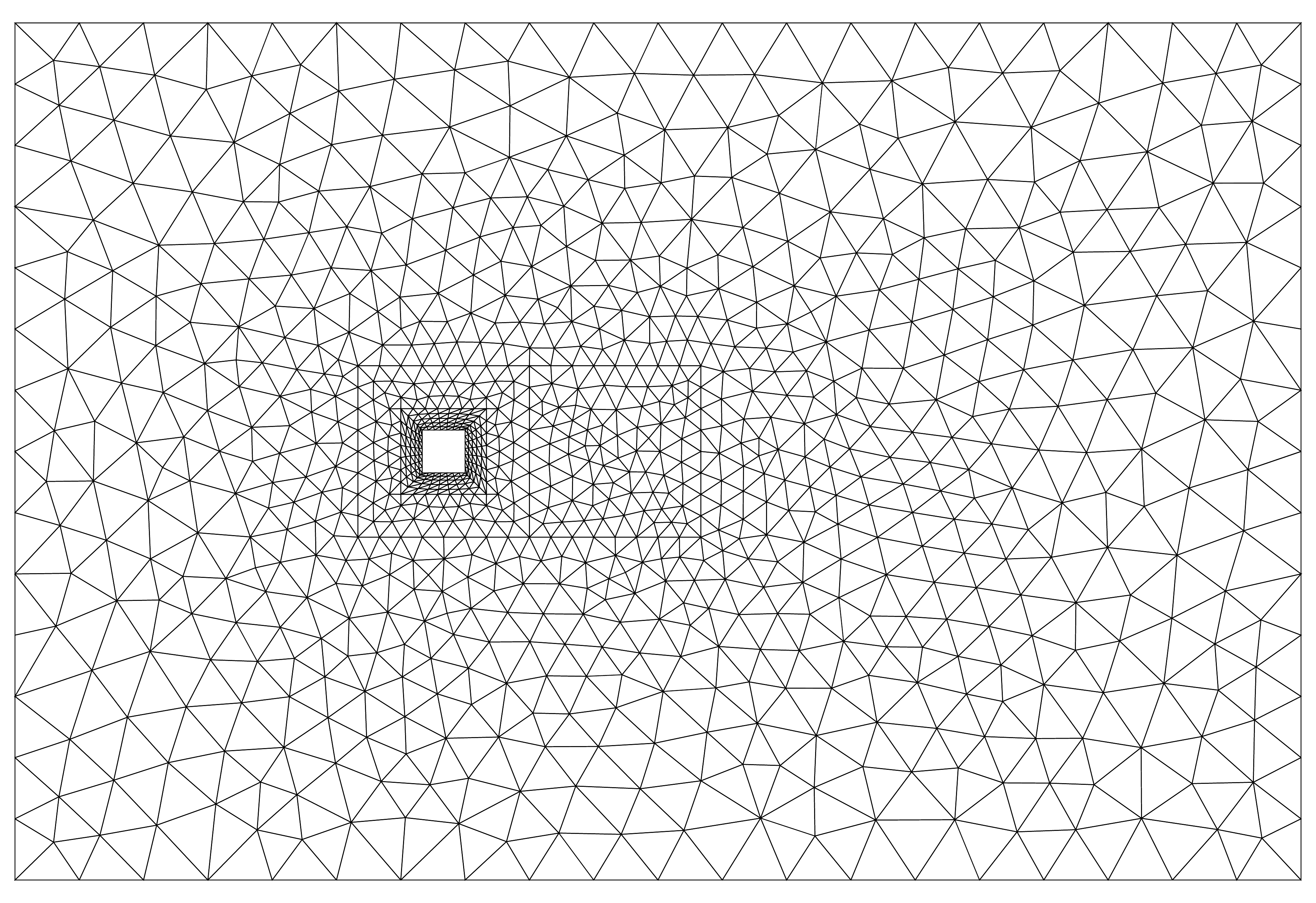}    
\caption{Two dimensional side view of the grid}
\label{fig:grid}
\end{figure*}

A sketch of the computational domain is shown in figure \ref{fig:cyl}. Given 
the square side $H,$ the length of the inflow part has been taken as 
$Lf = 10H$, the outflow $Lr = 20 H$, the upper and lower height $Ls = 10H$ and 
finally the geometry is extruded in the $z$ spanwise direction of $Lz = 4H$. 
The blockage of the cylinder is $5\%$, less  than that in the original experiment reported in
\cite{lyn:1995}.
While some authors \cite{liu:2012}, \cite{murakami:1999} tested different methods to 
enforce a certain level of turbulence at the inflow to recreate the conditions 
at the inflow of the experiment \cite{lyn:1995}, we decided for simplicity to 
use a uniform inflow condition, as done by other authors in different 
contexts, e.g.  \cite{sohankar:2000}, \cite{trias:2015}, \cite{verstappen:1998}.
Wall adhesion conditions are imposed on the cylinder and Dirichlet conditions 
with far field values on the inflow and outflow sides, with sponge layers to 
dampen the disturbances and avoid reflections, as discussed for example in  
 \cite{crivellini:2016}, \cite{restelli:2009}. 
On the upper and lower sides, Neumann boundary conditions are imposed, while in the 
spanwise direction periodic boundary conditions are enforced. 
The mesh used has 23816 tetrahedra and uses a structured block around
the cylinder, to better control the element anisotropy, while being
fully unstructured in the rest of the domain; this allows obtain an
adequate resolution on the cylinder sides, without an excessive number
of elements in 
the far field, as would happen with a fully extruded mesh. A two-dimensional 
side view of the mesh is presented  in figure \ref{fig:grid}.

Starting from a uniform initial condition, the simulation is evolved until statistical 
steady state is achieved. The resulting fields are used as a starting 
condition for the  simulations described in the following, in which statistics are accumulated 
for around 16 shedding periods,    averaging first along the spanwise 
direction and then in time.  After this time  span, which is indeed
longer than the averaging time prescribed in 
\cite{voke:1997}, flow statistics showed only minor  modifications.
Statistics will be then presented as plots along three lines in the $x-y$ plane, as illustrated in figure \ref{fig:cyl}.

\subsection{Constant degree simulations}
\label{subs:Constant degree simulations}
Some standard non adaptive simulations have been performed, to obtain a 
reference for the adaptive simulations with respect to the  quality of results and the
required computational effort. Three different simulations have been performed on the 
same grid with uniform polynomial degree equal to 2, 3 and 4. The grid has been 
dimensioned to be comparable with other similar computations 
\cite{minguez:2011}, \cite{sohankar:2000} using 
polynomial degree 4, thus the simulation at lower degree are considerably 
under-resolved. The subgrid model used for all the square cylinder 
computations is the Smagorinsky model described in section \ref{sec:dgles}. 
While more refined subgrid models have been already tested in the present 
computational framework, see e.g. \cite{abba:2015}, our goal here was to  
carry out the  analysis
of adaptive approaches   with the simplest possible subgrid model. 
The impact of more sophisticated models is the subject of ongoing work.

The global results, as well as computational data are presented in 
table~\ref{tab:ref-global}.
The Strouhal number is generally over-predicted with respect to the 
experimental values, as well as the drag coefficient and fluctuations. 
Some of these trends are reported also by other authors performing LES 
simulations, see e.g.  \cite{liu:2012}, \cite{minguez:2011}, \cite{matthieu:2008},  \cite{murakami:1999}, 
while the general over prediction of force coefficients might be due to the 
low compressibility effects introduced here but not in any of the other 
references. 

\begin{table*}
\centering
\begin{tabular}{rccccccc}
\toprule
            & St            & <Cd> & rms(Cd')    & rms(Cl')  & dofs & core h.    \\ \midrule
experiments & $\approx0.13$ & $\approx2.1$ & $\approx0.18$ & $\approx1.2$ & - & -\\
degree 2    & 0.1462 & 2.321 & 0.148  & 1.204 & 238160 & 785  \\
degree 3    & 0.1487 & 2.352 & 0.1605 & 1.305 & 476320 & 2665 \\
degree 4    & 0.1410 & 2.398 & 0.1930 & 1.374 & 833560 & 7561 \\
\bottomrule
\end{tabular}
\caption{Experimental and reference values of global quantities}
\label{tab:ref-global}
\end{table*}

\begin{figure*}
\centering
\begin{subfigure}[streamwise mean velocity]{\label{fig:ref_h1_umean}
    \includegraphics[width=0.45\textwidth]{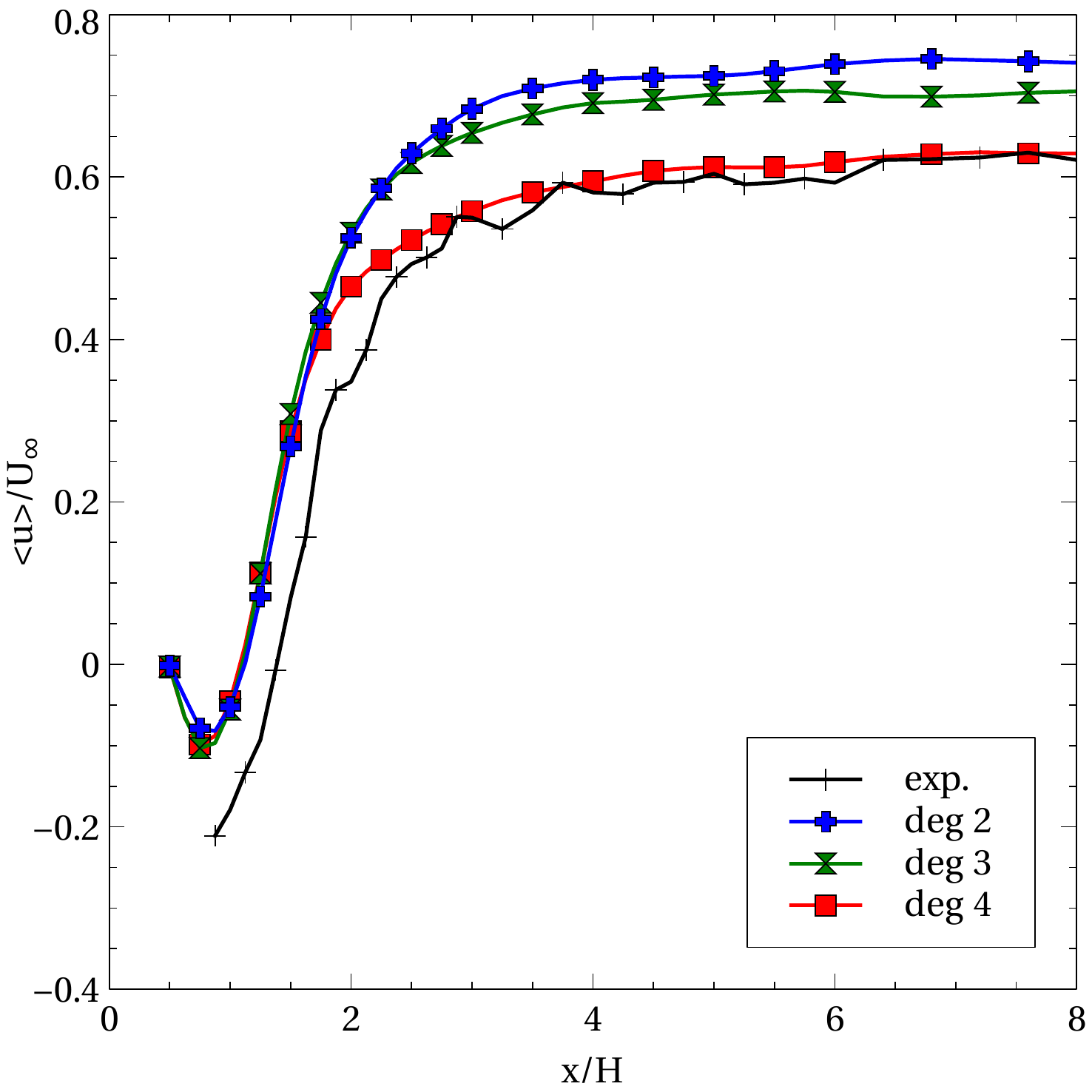}}   
\end{subfigure} \\
\begin{subfigure}[square root of total turbulent stresses, xx component]{\label{fig:ref_h1_rmsu}
    \includegraphics[width=0.45\textwidth]{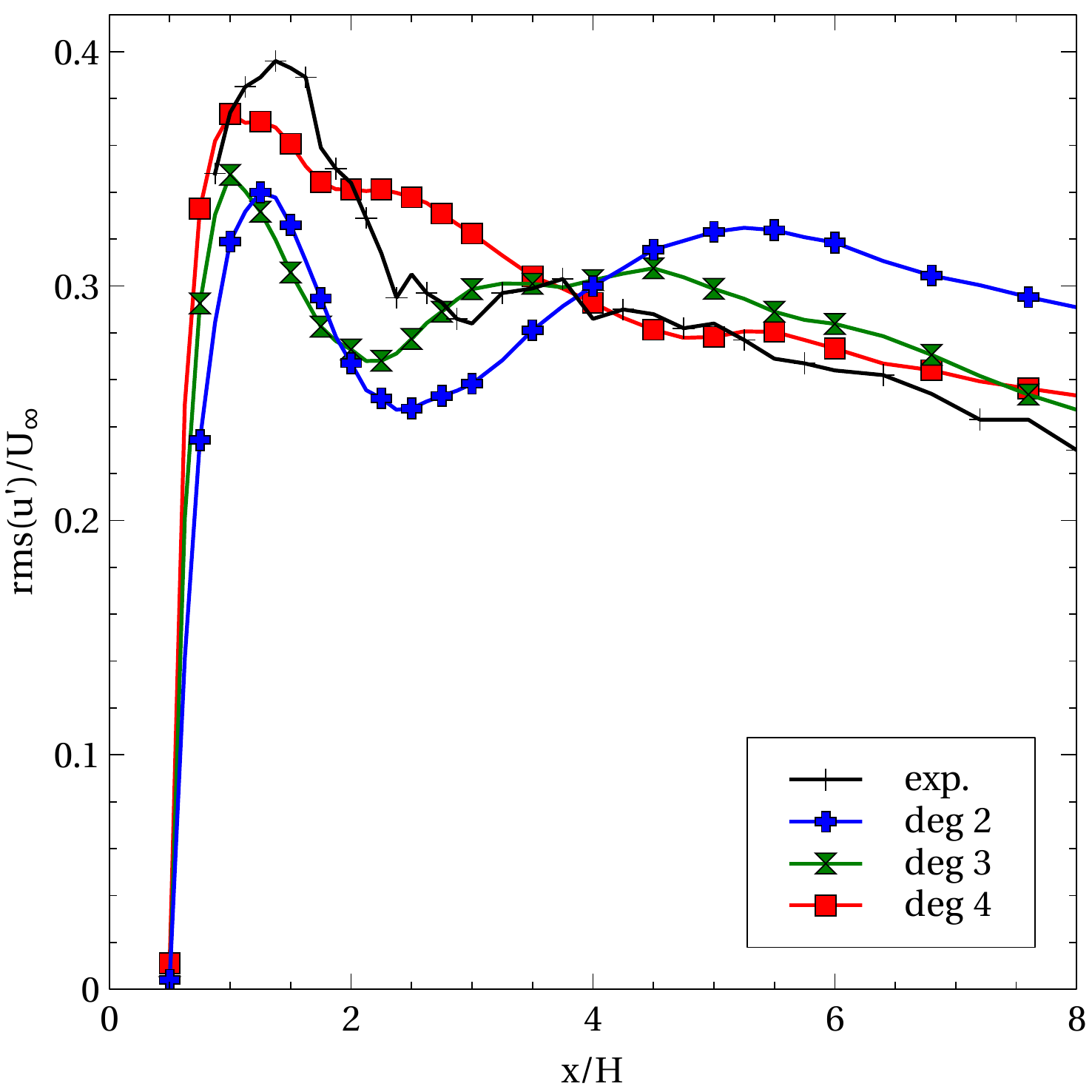}}    
\end{subfigure} 
\hfill
\begin{subfigure}[square root of total turbulent stresses, yy component]{\label{fig:ref_h1_rmsv}
    \includegraphics[width=0.45\textwidth]{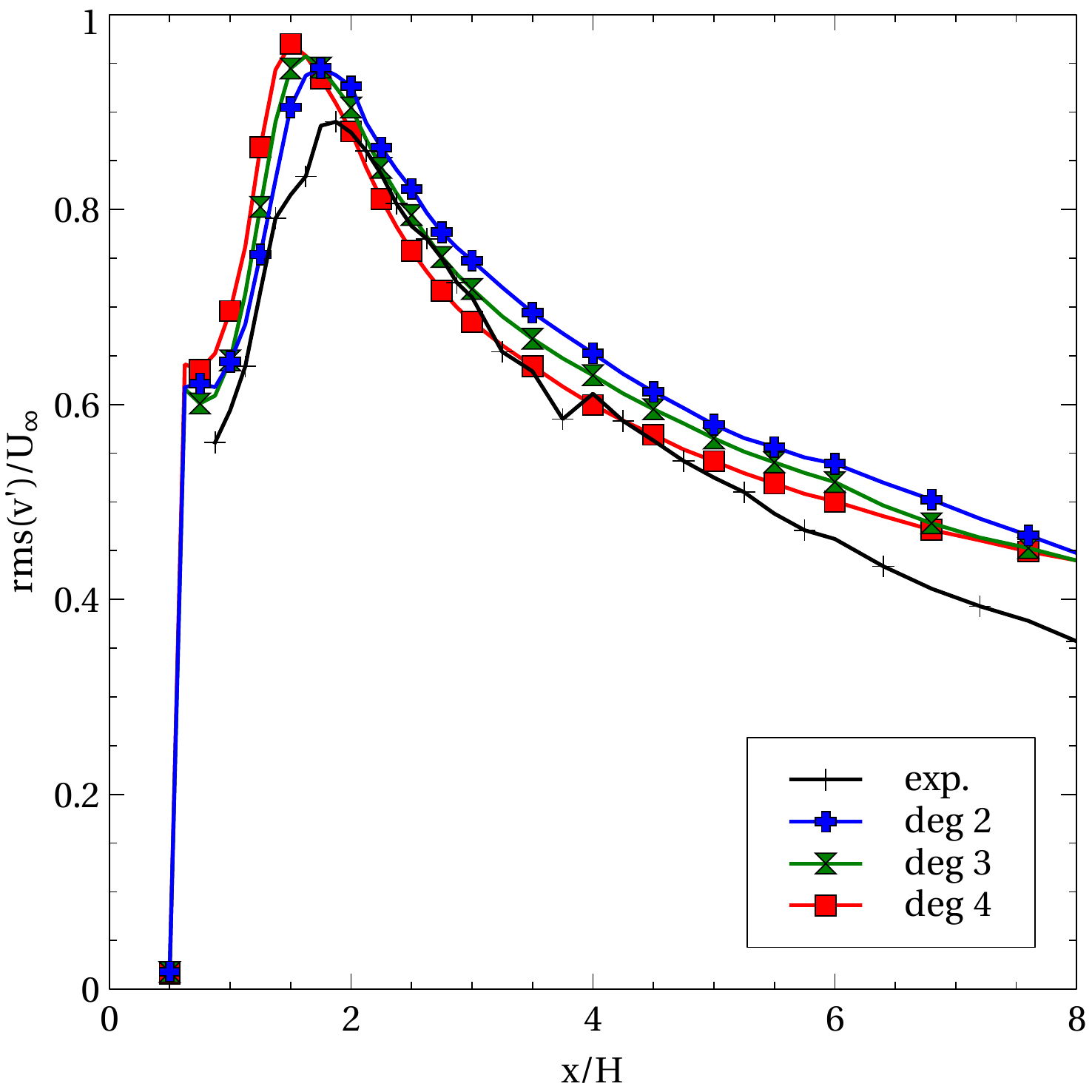}}    
\end{subfigure} 
\caption{Velocity statistics in the wake of the cylinder along plot line Z (see fig. \ref{fig:cyl}), comparison of the results with different uniform polynomial degree}
\label{fig:ref_h1}
\end{figure*}

\begin{figure*}
\centering
\begin{subfigure}[Streamwise mean velocity]{\label{fig:ref_v5_umean}
    \includegraphics[width=0.45\textwidth]{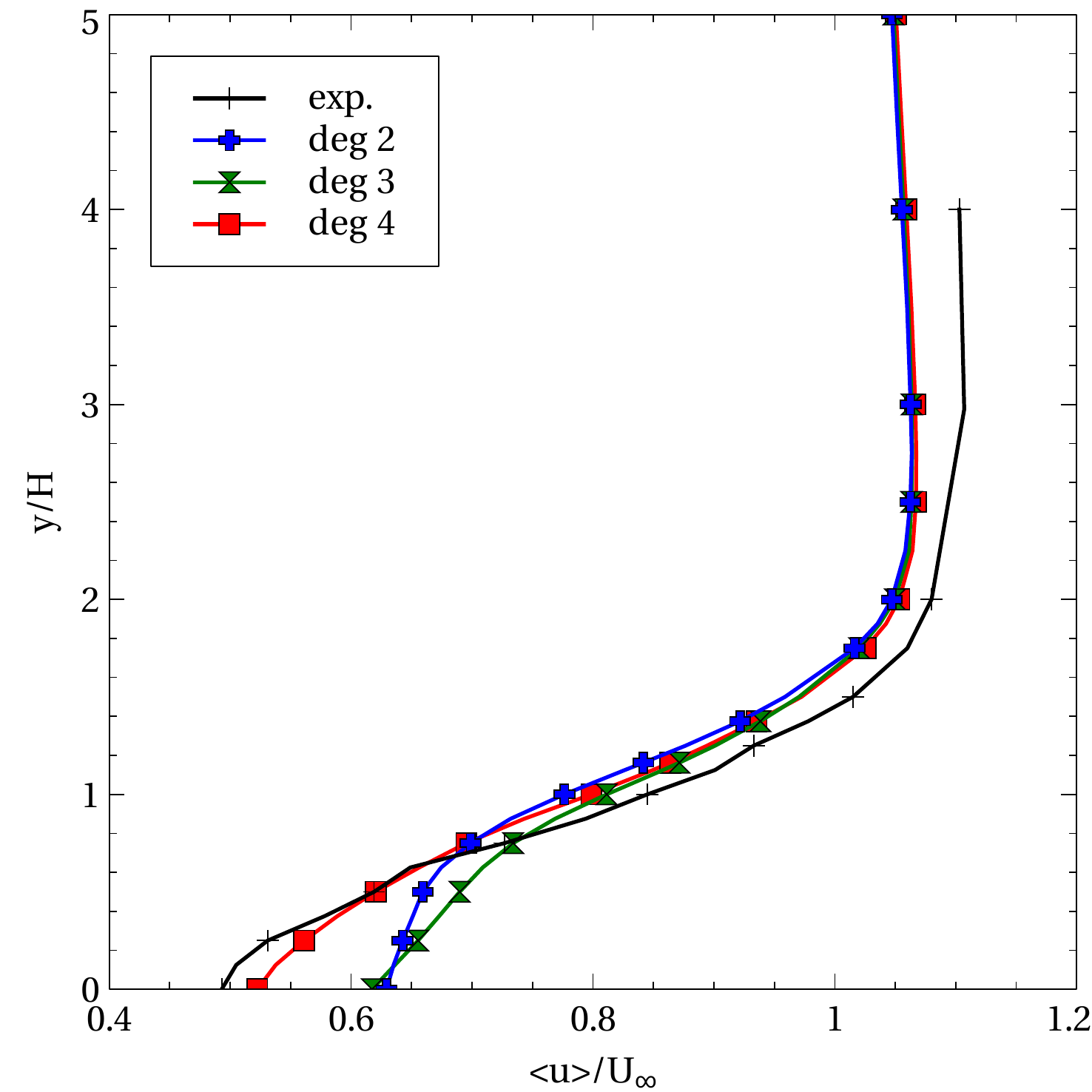}}    
\end{subfigure}\hfill 
\begin{subfigure}[total turbulent stresses, xy component]{\label{fig:ref_v5_tauuv}
    \includegraphics[width=0.45\textwidth]{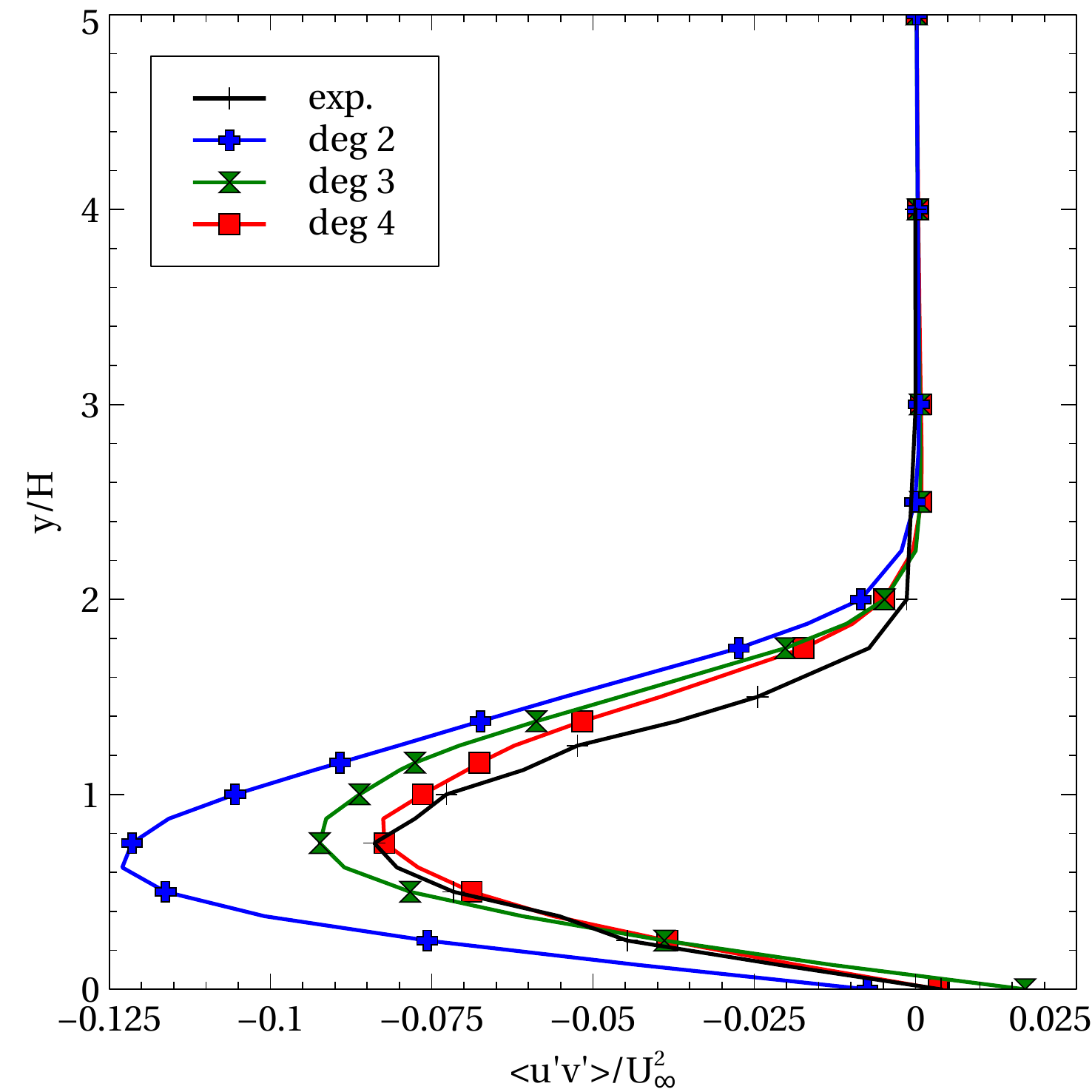}}    
\end{subfigure}\\
\begin{subfigure}[square root of total turbulent stresses, xx component]{\label{fig:ref_v5_rmsu}
    \includegraphics[width=0.45\textwidth]{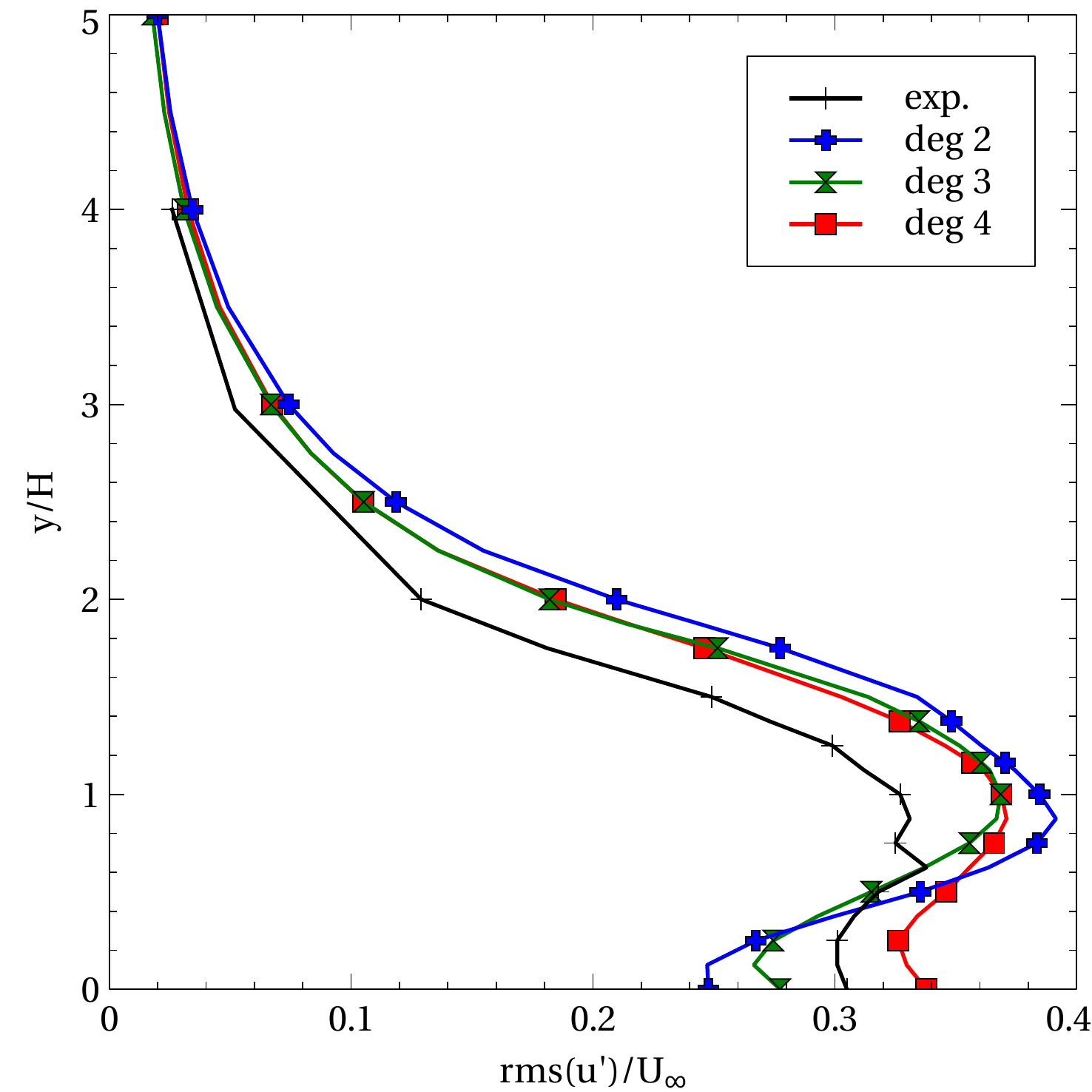}}    
\end{subfigure} 
\hfill
\begin{subfigure}[square root of total turbulent stresses, yy component]{\label{fig:ref_v5_rmsv}
    \includegraphics[width=0.45\textwidth]{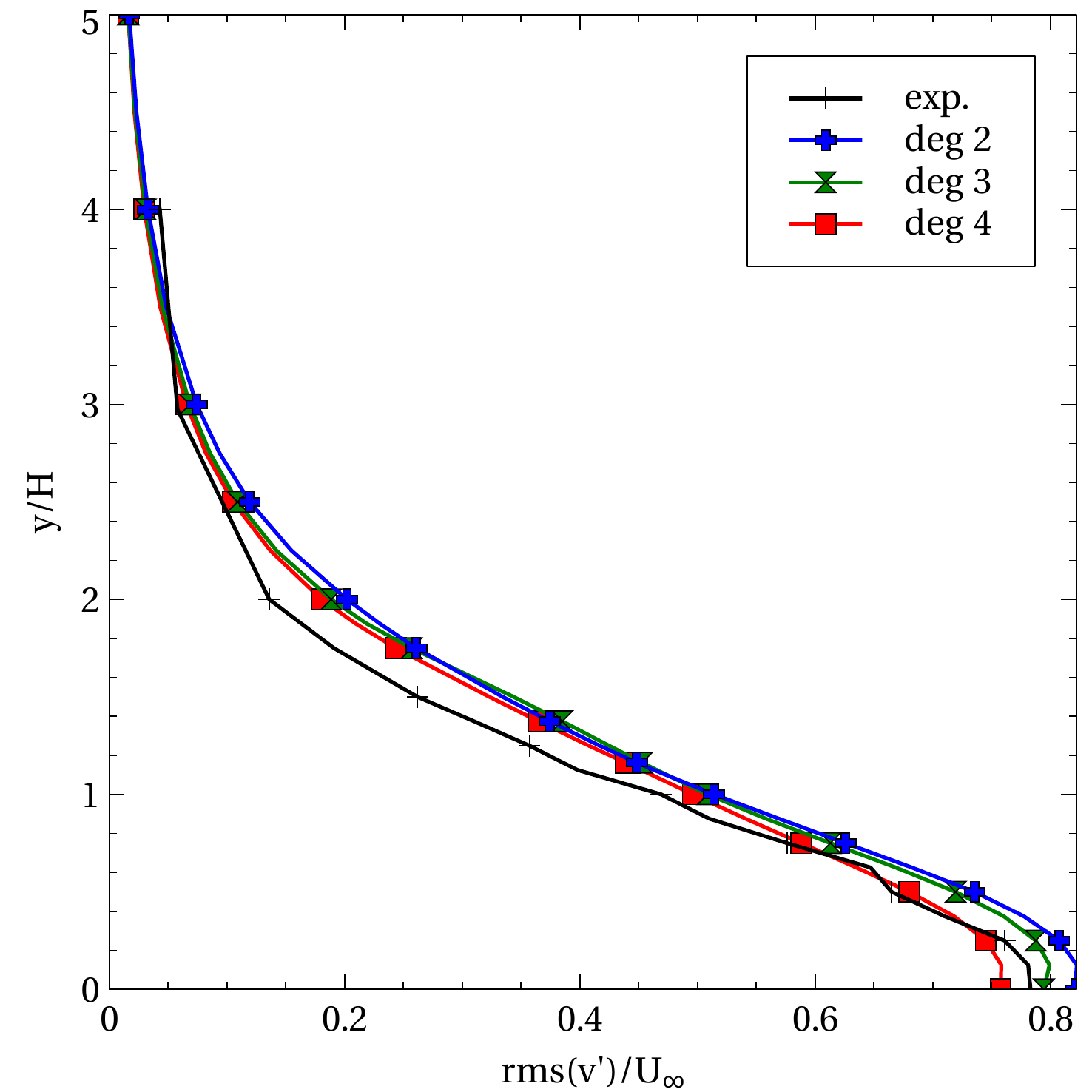}}    
\end{subfigure} 
\caption{Velocity statistics across the wake of the cylinder, along plot line W (see fig. \ref{fig:cyl}), comparison of the results with different uniform polynomial degree}
\label{fig:ref_v5}
\end{figure*}

\begin{figure*}
\centering
\begin{subfigure}[Streamwise mean velocity]{\label{fig:ref_v3_umean}
    \includegraphics[width=0.22\textwidth]{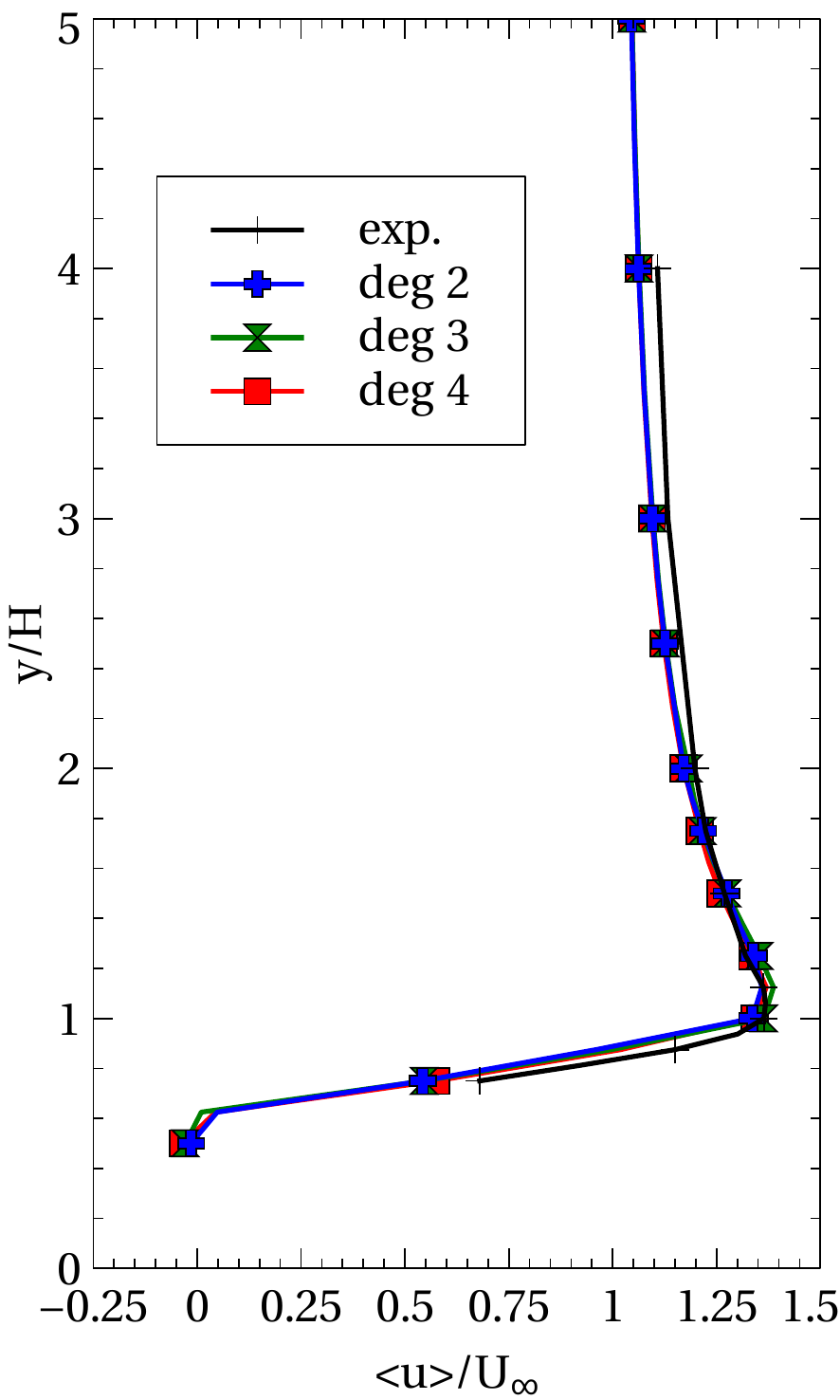}}    
\end{subfigure}
\begin{subfigure}[square root of total turbulent stresses, xx component]{\label{fig:ref_v3_rmsu}
    \includegraphics[width=0.22\textwidth]{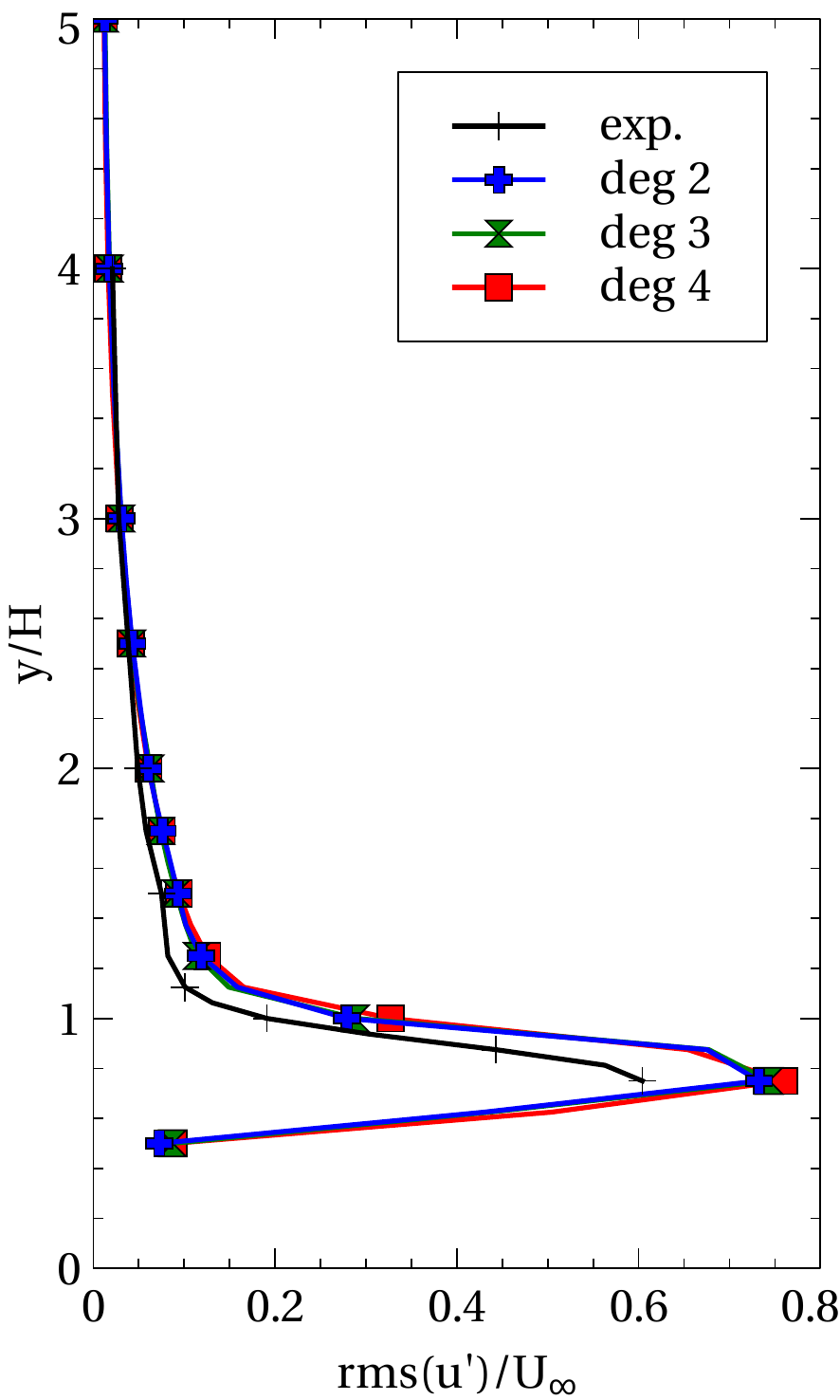}}    
\end{subfigure} 
\begin{subfigure}[square root of total turbulent stresses, yy component]{\label{fig:ref_v3_rmsv}
    \includegraphics[width=0.22\textwidth]{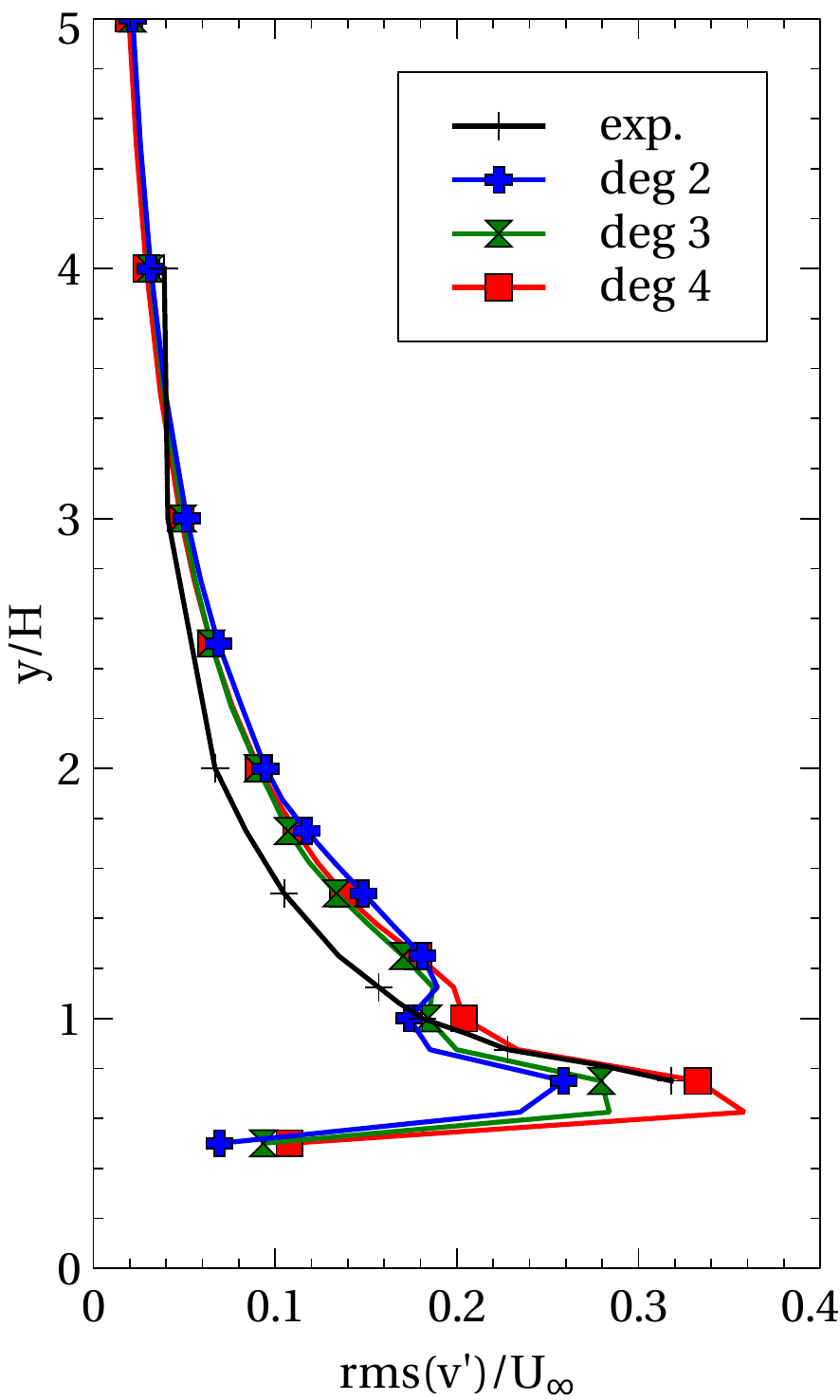}}    
\end{subfigure} 
\begin{subfigure}[total turbulent stresses, xy component]{\label{fig:ref_v3_tauuv}
    \includegraphics[width=0.22\textwidth]{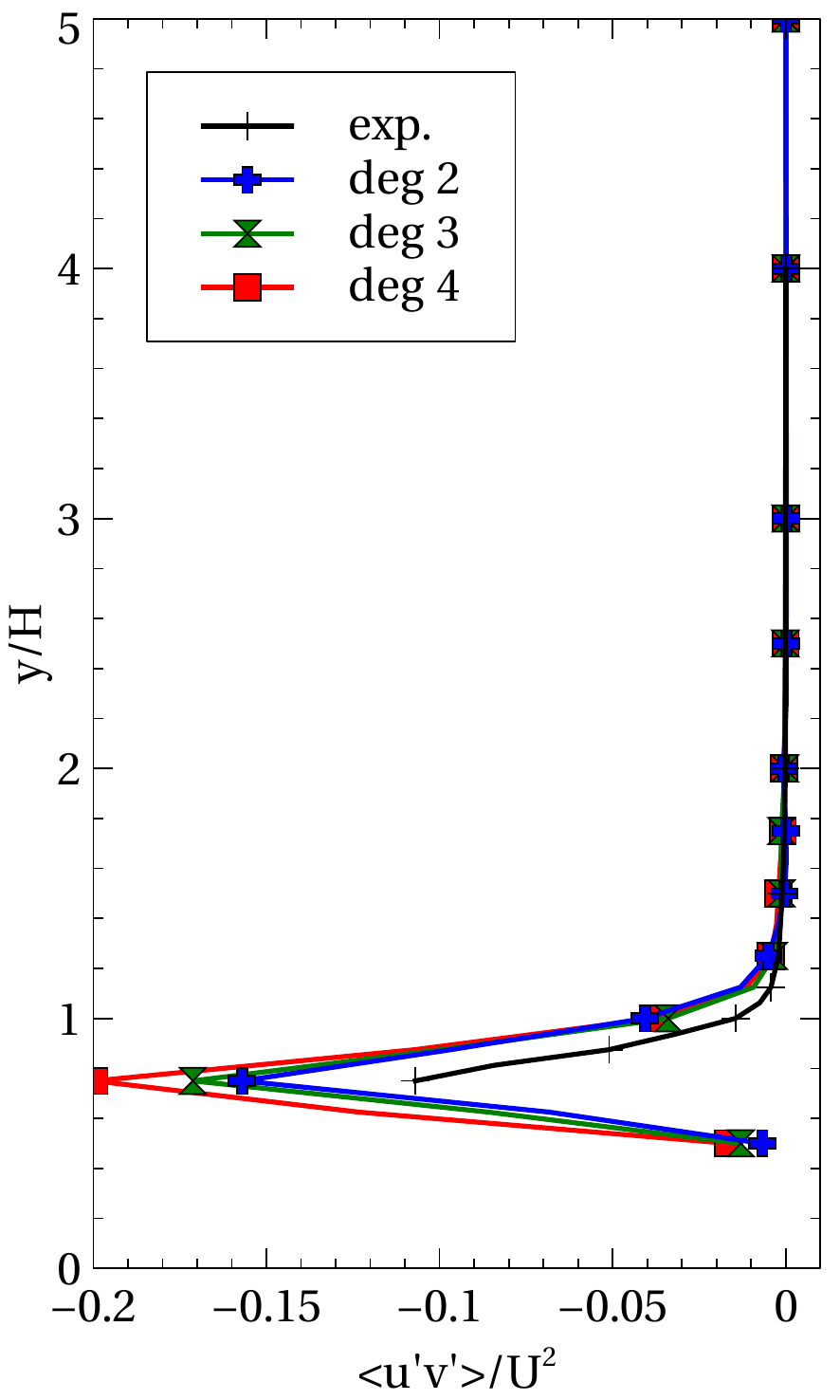}}    
\end{subfigure}
\caption{Velocity statistics over the cylinder in correspondence of the 
rear corner, along plot line K (see fig. \ref{fig:cyl}), comparison of the results with different uniform polynomial degree}
\label{fig:ref_v3}
\end{figure*}

The profiles of some velocity statistics and total turbulent stresses 
(resolved and modelled) are presented in 
figures~\ref{fig:ref_h1},~\ref{fig:ref_v5} and~\ref{fig:ref_v3}, measured along 
a line in the wake center, a line across the wake two diameters downstream and a 
vertical line over the rear corner of the cylinder.  First of all,  when increasing the polynomial degree,
we   observe a substantial convergence of the numerical solution to 
the experimental values  in the wake, while 
over the rear corner of the cylinder the results are closer for all the different polynomial 
orders employed. The solutions computed with degree 2 
are unsatisfactory in many areas, but more interestingly the solutions  with degree 3, 
while being generally better than the degree 2 ones, are not a substantial 
improvement in the wake, where only the discretization with degree 4 gives acceptable 
results. 

As noted above for the global results, these profiles
are in general agreement with the experimental results, 
but still there are some discrepancies, which may be caused by employing 
compressible equations, or the lack of turbulence at inlet.  The mismatch with
 experimental measurements is similar to the 
one observed in other computations, e.g. \cite{murakami:1999}, \cite{rodi:1997a},
\cite{sohankar:2000}. Notice that  even experimental records such as those reported 
in \cite{lyn:1995} show a lack of agreement 
with other experimental data beyond 3 diameters downstream in the wake.
The lower velocity far from the wake in figure \ref{fig:ref_v5_umean} is 
instead clearly caused by the lower blockage of the cylinder in the simulation 
with respect to the experiment. 

Since the simulations have been performed in different conditions than the 
experiments with which they have been are compared, some discrepancies, especially in the wake 
area, are justifiable and the results can be considered accurate enough. 
Therefore, for the following adaptive simulations, the main reference will be the 
refined constant degree simulations, rather than experimental results.


\subsection{Validation of the adaptive simulations}
\label{subs:adapt from ref}
After the reference solutions have been obtained,  adaptive simulations 
have been performed. 
The aim of the  adaptive computations presented in the following is to validate the 
adaptation procedure, to assess the quality of the results compared with the 
reference non adaptive results, to compare the performances of the proposed 
indicators and to check the reduction in computational effort with respect to the
non adaptive simulation. The adaptation 
procedure was based on indicators computed on preliminary runs that employed
constant polynomial degree equal to 4, thus thwarting the reduction in 
computational effort obtained with the adaptation.
This however allows us to validate the proposed approach, while
adaptive simulations based 
on cheaper adaptation procedure and yielding   a net reduction of the
 computational effort will  be presented in the following section \ref{subs:adapt from low deg}.

Both the   indicator based on modal coefficients (M) and the structure function (SF) 
indicator introduced in section \ref{sec:adapt} have been computed based on the 
results obtained from a preliminary simulation with constant degree 4.
Two threshold values  $\epsilon_1  $   $\epsilon_2  $  have been chosen for each indicator. 
The threshold values were selected to obtain a reasonable amount of degrees of 
freedom and an acceptable distribution of polynomial degrees, as well as to obtain a comparable 
number of degrees of freedom for
  the two adaptive simulations. More specifically, for the
 indicator based on modal coefficients (M)  the values
   $\epsilon_1 = 0.075, $   $\epsilon_2  =  0.113$ were used, while 
  for the structure function (SF) the values
   $\epsilon_1 = 5\times 10^{-4}, $   $\epsilon_2  =  1\times 10^{-2}$ were used.
The  cells with indicator values smaller than  $\epsilon_1  $ have been assigned polynomial degree 2, 
those with indicator values larger than  $\epsilon_2 $ have been assigned polynomial degree 4,
 while the others have been assigned polynomial degree 3.
The  adaptive simulations have been initialized with appropriate local projections
of the     previous uniform degree 4 results.

\begin{figure*}
\centering
\begin{subfigure}[M indicator]{\label{fig:degmap_rw2}
    \includegraphics[width=0.8\textwidth]{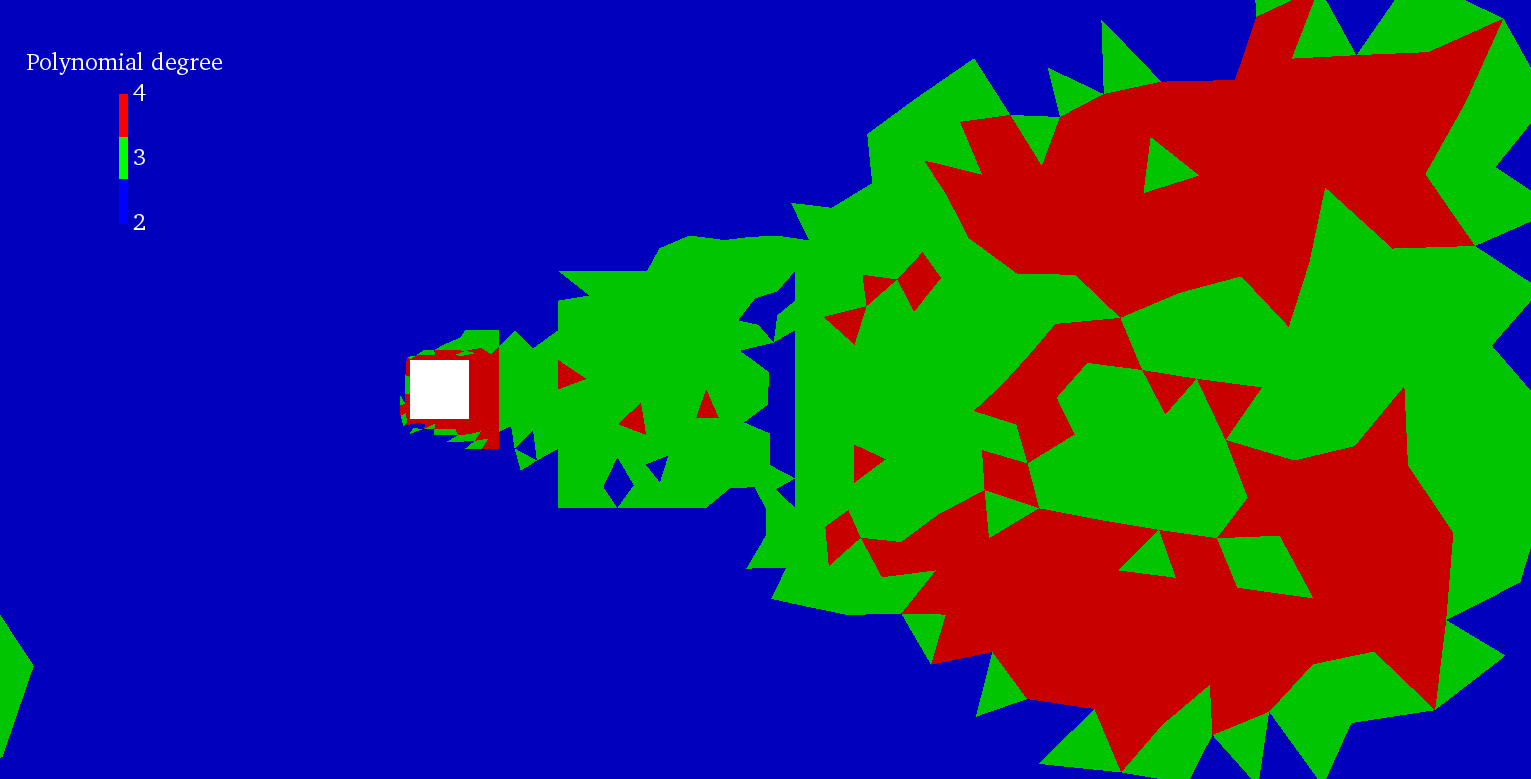}}    
\end{subfigure}\\
\begin{subfigure}[SF indicator]{\label{fig:degmap_sf1}
    \includegraphics[width=0.8\textwidth]{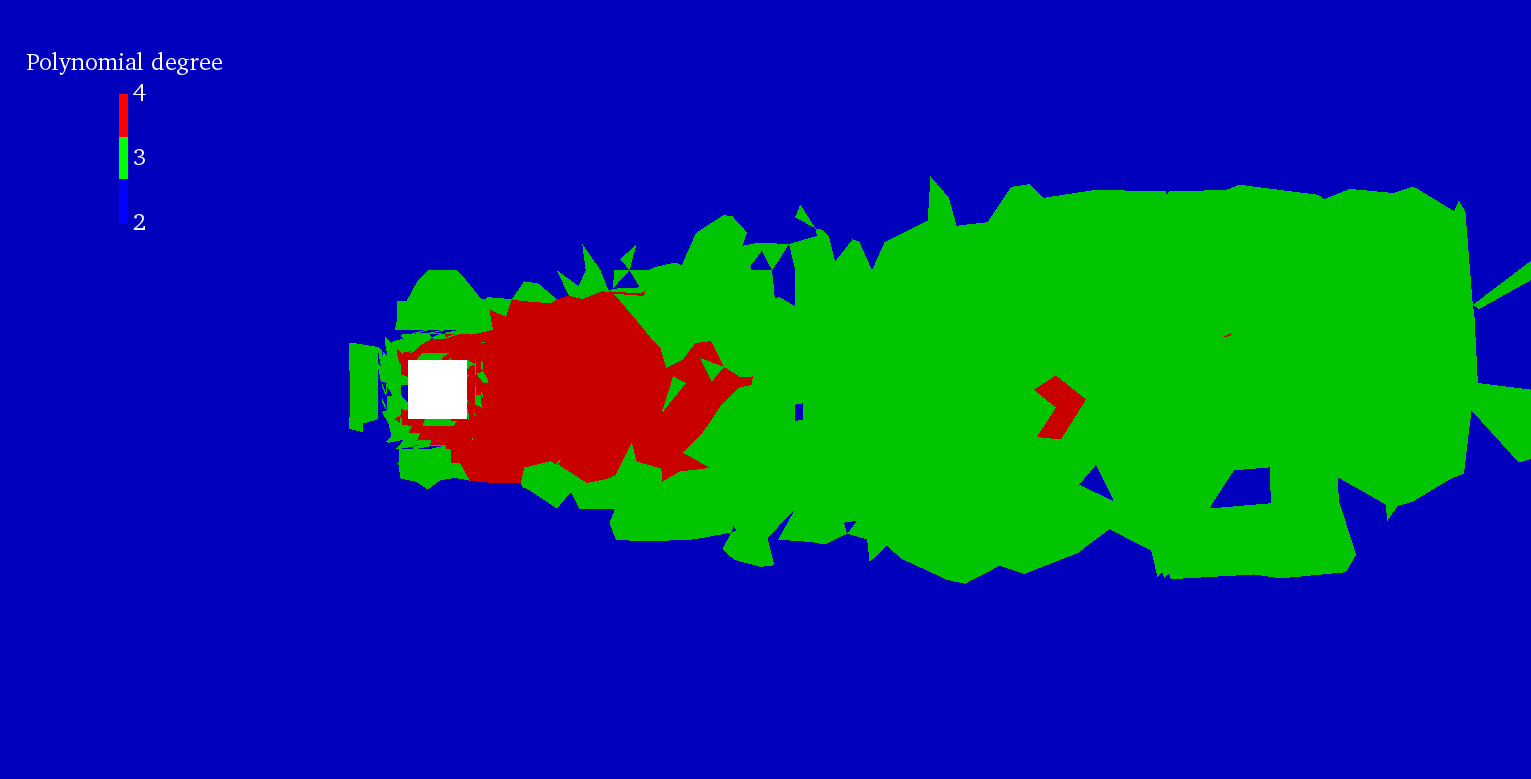}}    
\end{subfigure} 
\caption{Map of the polynomial degree of the elements, on a vertical slice, for the two different indicators}
\label{fig:degmap}
\end{figure*}

Both   refinement indicators generally show larger values in the wake of the 
cylinder and lower values away from it, as it is predictable and desirable. 
The M indicator yields a narrower range of values, from $O(10^{-4})$ to 
$O(10^{-1})$, with larger values  more scattered in the wake, meaning that 
the high order elements are distributed over a wider area. 
 The SF indicator yields instead a broader  value range, from $O(10^{-12})$ to 
$O(1)$, with the larger values concentrated in the near wake of the cylinder and in the shear layer generated by the separations after the front vertexes of the cylinder.
In the wake, the indicator takes intermediate values, while outside the 
wake it reaches very low values. This allows to identify more clearly the area to be
refined. 
The resulting polynomial distributions are shown in figure \ref{fig:degmap}, on 
a 2D plane perpendicular to the spanwise direction. Along the spanwise direction,
the degree distribution is 
almost uniform  and small differences are due to 
the fact that the mesh is not uniform in that direction. The polynomial 
distribution in case of the M indicator shows a narrow high order zone in 
close proximity of the cylinder wall, and a broader higher order area far in 
the wake of the cylinder. In case of the SF indicator, the high order area is 
mainly confined in the near wake of the cylinder and around the forward 
corners, and generally the more refined zone has approximately 
at the same width throughout the wake. 

The domain decomposition for adaptive parallel runs was performed using the METIS library \cite{karypis:1998fast}, as in the non adaptive ones. However, in order to ensure an efficient load balancing, each cell has been assigned a different weight in the graph partitioning algorithm, based on the element polynomial degree. Some experiments have been performed by using different powers of the number of degrees of freedom of the element,  as a weight, but in the end the simple  weighting by the number of element degrees of freedom resulted in the the best parallel load balance. 

\begin{table*}
\centering
\begin{tabular}{rccccccc}
\toprule 
                & St            & <Cd> & rms(Cd')    & rms(Cl')  & dofs & core h.    \\ \midrule
degree 4        & 0.1410      & 2.398           & 0.1930      & 1.374 & 833560 &  7561   \\
adaptive M        & 0.1594      & 2.293           & 0.1319      & 1.191 & 398160 & 2949  \\
adaptive SF        & 0.1410      & 2.350           & 0.1692      & 1.241 & 398270 &  2662    \\
\bottomrule
\end{tabular}
\caption{Global quantities in  $p-$adaptive simulations}
\label{tab:adapt-global}
\end{table*}

The global data of the adaptive simulations are presented in table \ref{tab:adapt-global}, compared with the  uniform degree 4 simulation. 
It is clear that the values obtained from the simulation with the SF indicator are closer to those of the  full degree 4 simulation than those obtained from the simulation performed with the M indicator. 
With respect to the full degree 4 simulation, there is a reduction of $52\%$ of the degrees of freedom from both simulations and a mean reduction of the computational time of around $60\%$. 
Notice
that, when considering $p-$adaptation, the computational cost is not
directly proportional to the number of degrees of freedom, since
within a high order element more degrees of freedom are coupled with
each other. Hence, reducing the number of high order elements has a
beneficial effect that goes beyond the mere reduction of the total
number of degrees of freedom.
Fluctuations in the computational time are due to possible unbalances during parallel runs. 

\begin{figure*}
\centering
\begin{subfigure}[streamwise mean velocity]{\label{fig:adapt_h1_umean}
    \includegraphics[width=0.45\textwidth]{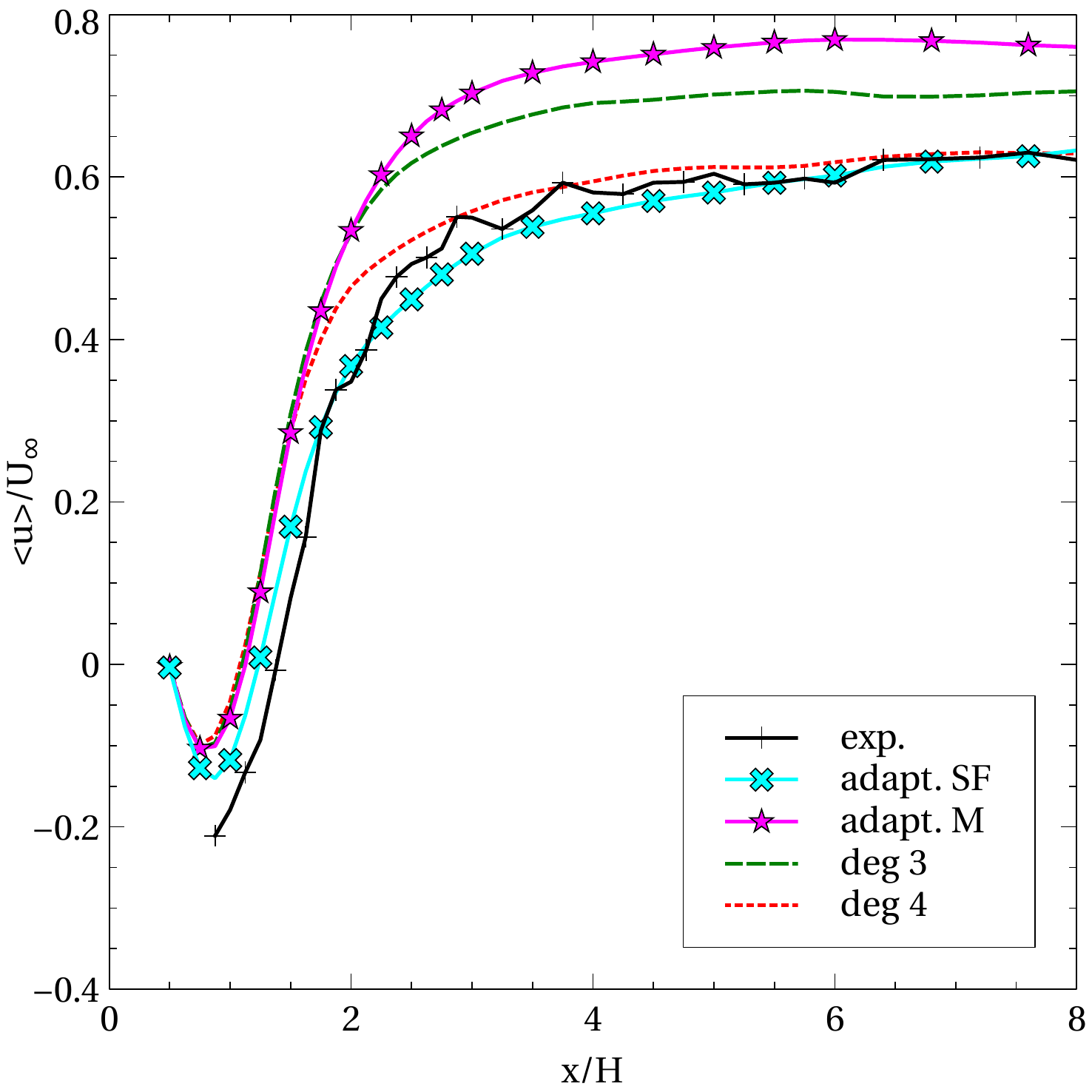}}   
\end{subfigure} \\
\begin{subfigure}[square root of total turbulent stresses, xx component]{\label{fig:adapt_h1_rmsu}
    \includegraphics[width=0.45\textwidth]{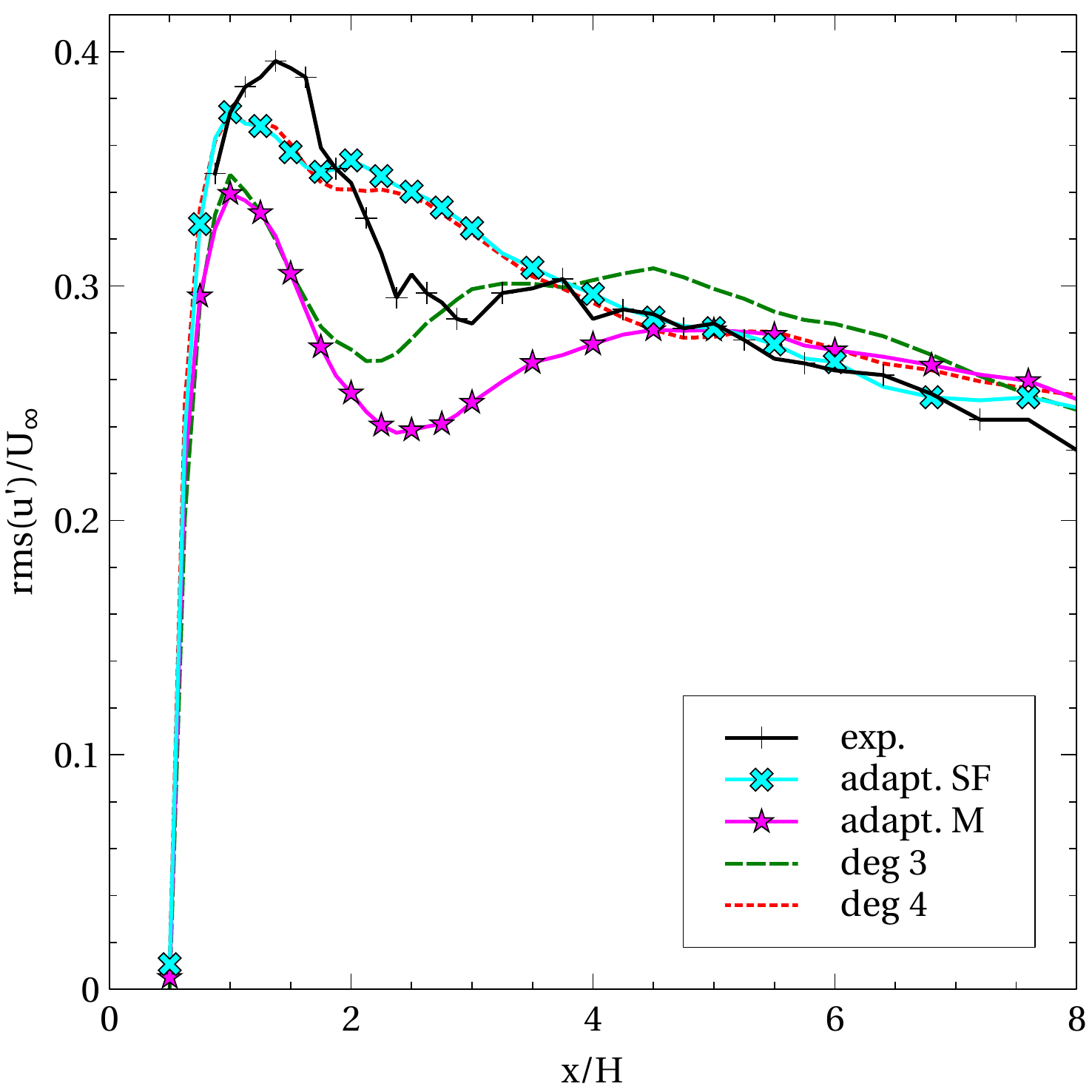}}    
\end{subfigure} 
\hfill
\begin{subfigure}[square root of total turbulent stresses, yy component]{\label{fig:adapt_h1_rmsv}
    \includegraphics[width=0.45\textwidth]{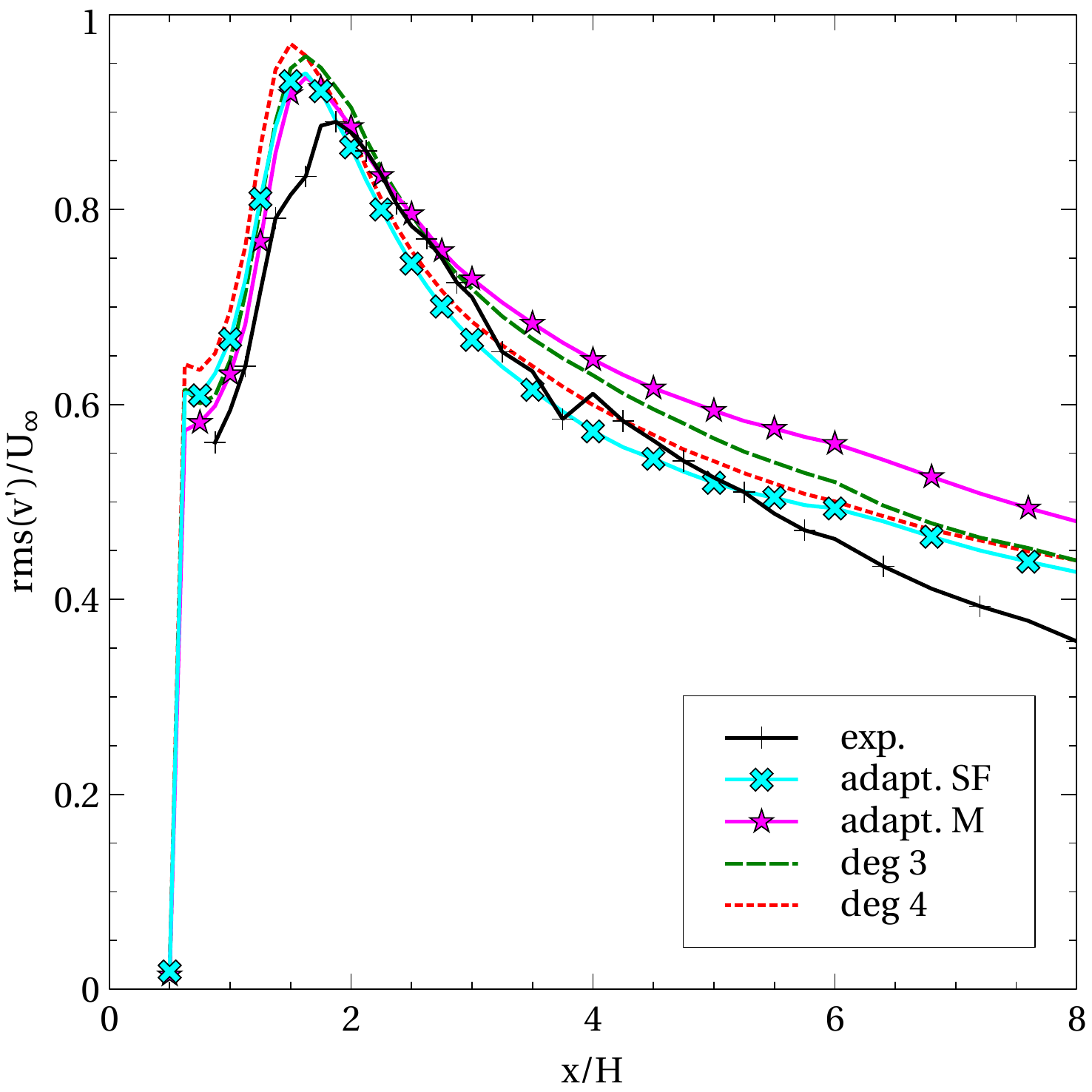}}    
\end{subfigure} 
\caption{Velocity statistics in the wake of the cylinder, along plot line Z (see fig. \ref{fig:cyl}), adaptive solutions compared with uniform degree solutions and experimental data}
\label{fig:adapt_h1}
\end{figure*}

\begin{figure*}
\centering
\begin{subfigure}[Streamwise mean velocity]{\label{fig:adapt_v5_umean}
    \includegraphics[width=0.45\textwidth]{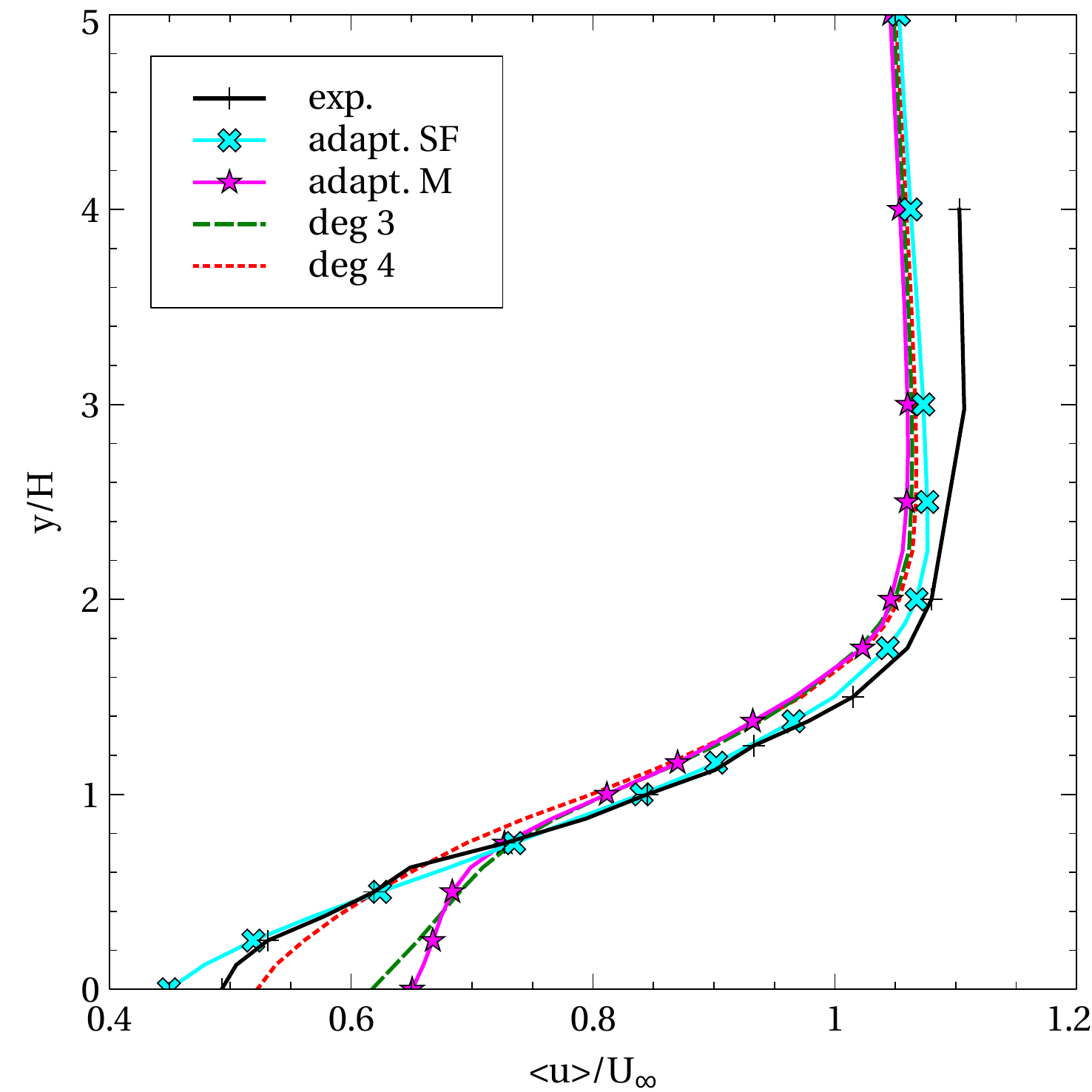}}    
\end{subfigure}\hfill 
\begin{subfigure}[total turbulent stresses, xy component]{\label{fig:adapt_v5_tauuv}
    \includegraphics[width=0.45\textwidth]{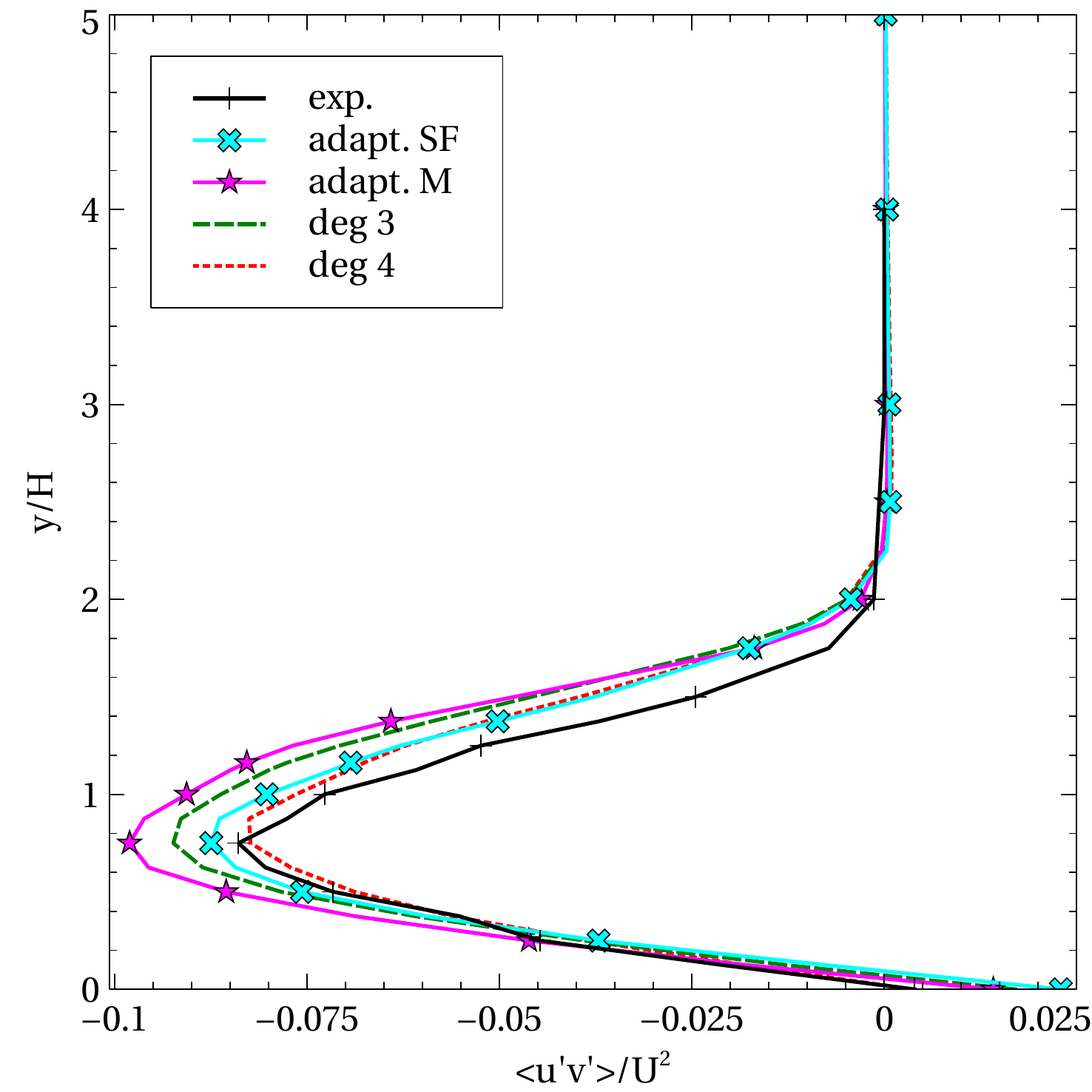}}    
\end{subfigure}\\
\begin{subfigure}[square root of total turbulent stresses, xx component]{\label{fig:adapt_v5_rmsu}
    \includegraphics[width=0.45\textwidth]{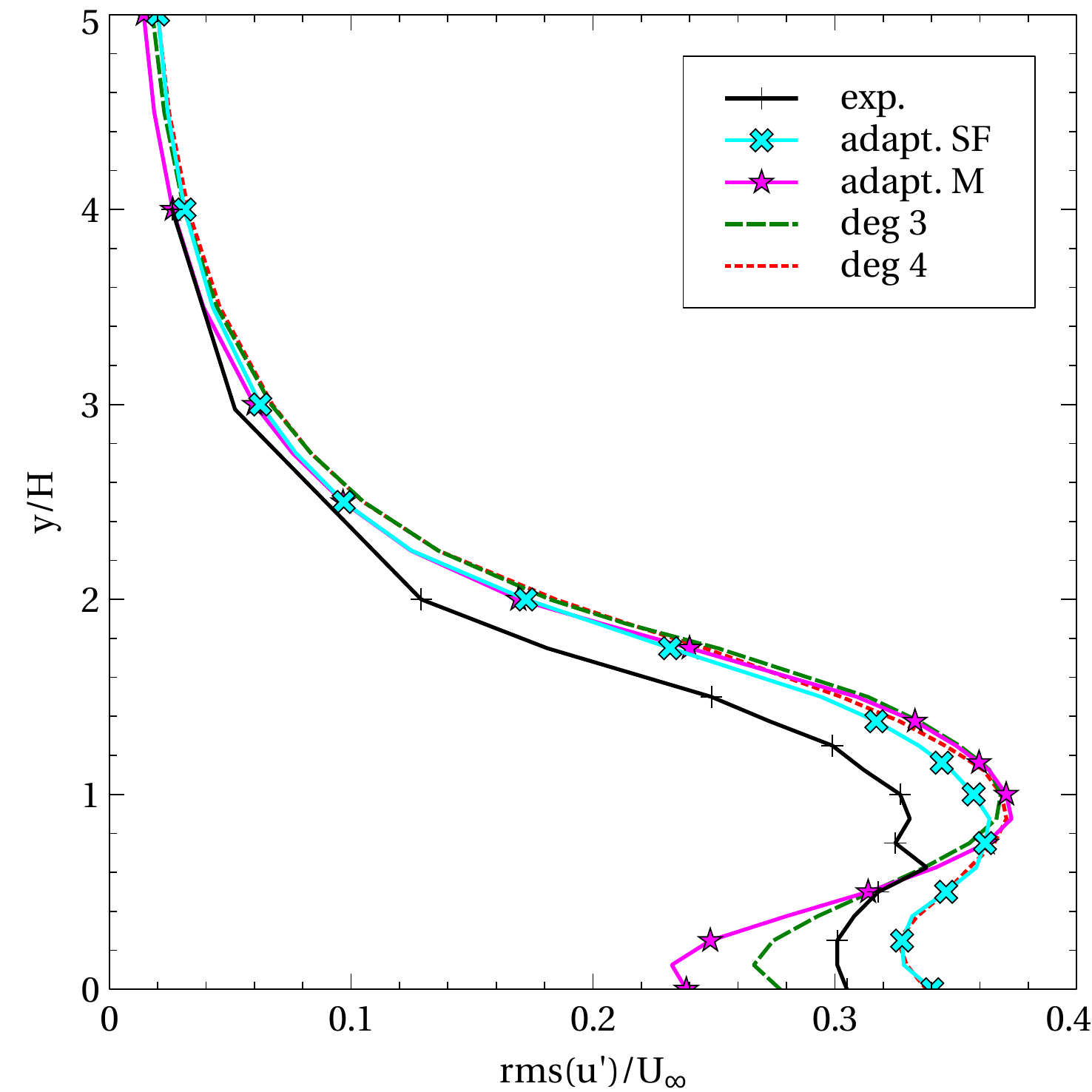}}    
\end{subfigure} 
\hfill
\begin{subfigure}[square root of total turbulent stresses, yy component]{\label{fig:adapt_v5_rmsv}
    \includegraphics[width=0.45\textwidth]{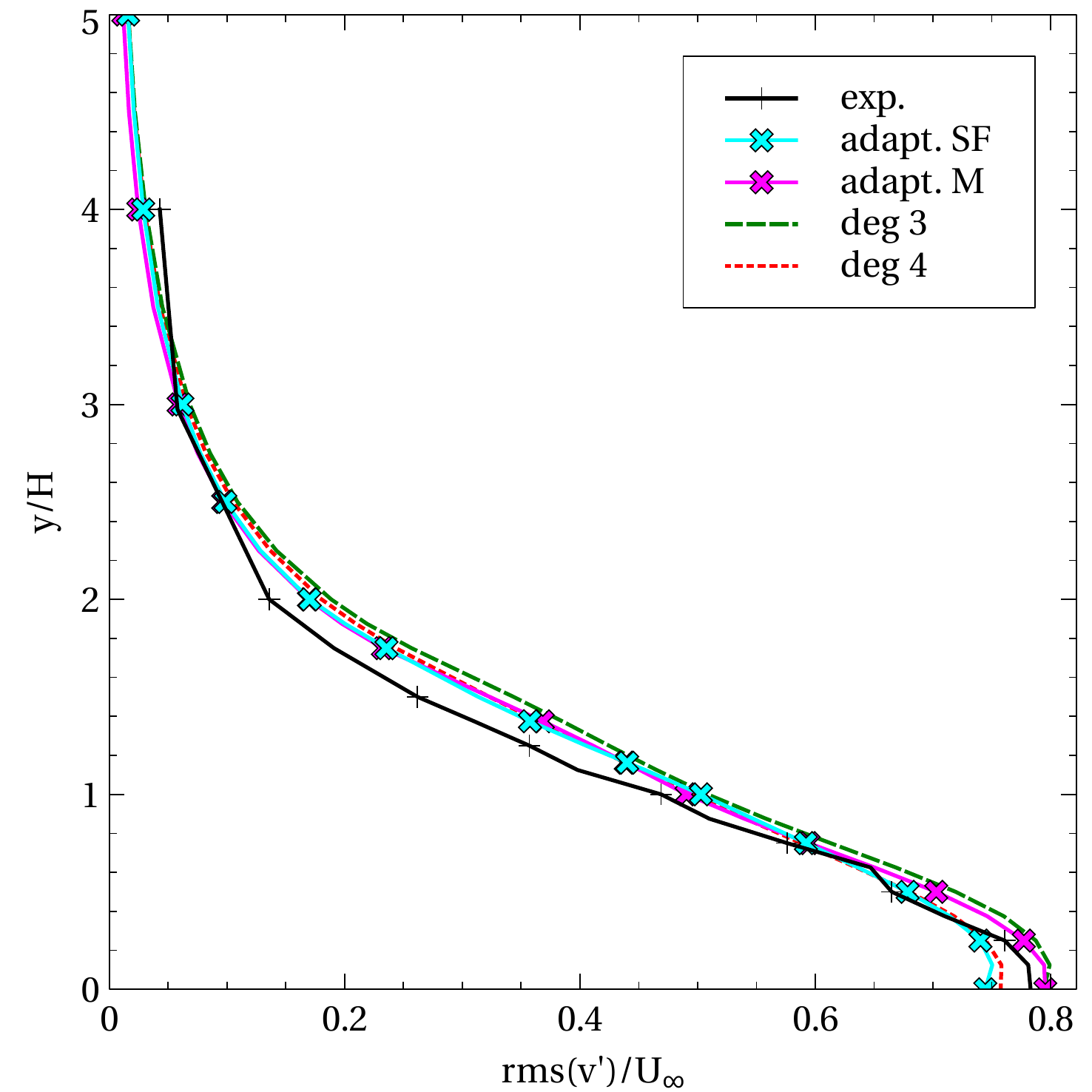}}    
\end{subfigure} 
\caption{Velocity statistics across the wake of the cylinder, along plot line W (see fig. \ref{fig:cyl}), adaptive solutions compared with uniform degree solutions and experimental data}
\label{fig:adapt_v5}
\end{figure*}

The profiles of the  velocity statistics are presented in figures \ref{fig:adapt_h1} and \ref{fig:adapt_v5}. It is evident that in most  cases the results obtained with the SF indicator are much closer to the reference results, while using the M indicator leads, with the same number of degrees of freedom, to results closer to those obtained with uniform degree 3, which as discussed previously are not   satisfactory. It is worth noting that, for both  adaptive simulations, the number of degrees of freedom is slightly smaller than that in the full degree 3 simulation, while the computational time is comparable. Summarizing, adapting the solution according to the M indicator allows to have  results similar to those of  the uniform degree 3 simulations with similar effort, i.e. it does not affect negatively the results, but  only yields a small improvement. On the other hand,   using the SF indicator it is possible to obtain results comparable to those of a degree 4 solution with the effort of a degree 3 
solution.


Given the presented results, the SF indicator appears  to be more robust and to lead to more accurate results. In this context of static adaptivity,  where the indicator calculation is performed as a local postprocessing task between a initial run and the actual adaptive run, the additional computational cost due to the more complex indicator is negligible. During the static indicator computation, only a few hundreds of full instantaneous fields from a previous run are processed and the impact of the indicator calculation is limited. In fact, all indicators calculations did not exceed a few minutes of CPU time on a personal computer. 
The cost of the indicator calculation will be  surely larger in  future dynamically adaptative runs, where the need to have a time resolved adaptation and a statistically reliable indicator evaluation will lead to more frequent evaluations of the indicator. 

Considering the results with the perspective of adaptation based on the physical local conditions of the flow, as discussed in sections \ref{sec:intro} and \ref{sec:adapt},  it is possible to correlate the better results obtained with the SF indicator to the fact that it allows to have a better physical insight on the actual flow conditions than the M indicator. This encourages   pursuing the study of other possible adaptation criteria based on the evaluation of the local flow properties. 

\subsection{Sensitivity of the adaptive simulation on the resolution of the indicator computation}
\label{subs:adapt from low deg}

Some simple sensitivity tests have been performed with respect to the data used to calculate the indicator and  the resulting polynomial distributions.
While keeping the rest of the procedure, thresholds included, identical to that described in section \ref{subs:adapt from ref},
we have used as data for the indicator calculation the results from 
\begin{itemize}
\item [1] a different degree 4 simulation with identical setup as the one presented in section \ref{subs:Constant degree simulations} 
\item [2] the results obtained with  the uniform degree 2 simulation presented in section \ref{subs:Constant degree simulations}. 
\end{itemize}
 In the first case the aim is to test the variability of the final results of the adaptivity procedure starting from different realizations of the same simulation. In the second, we wanted to investigate whether it was possible to use the much less expensive degree 2 results to calculate the indicator, thus obtaining a real reduction in computational time with respect to the uniform degree simulations. 
In both   cases, the polynomial distribution obtained is close to the one obtained from degree 4, first realization, shown in figure \ref{fig:degmap_sf1}.

\begin{figure*}
\centering
\begin{subfigure}[streamwise mean velocity]{\label{fig:adaptlow_h1_umean}
    \includegraphics[width=0.45\textwidth]{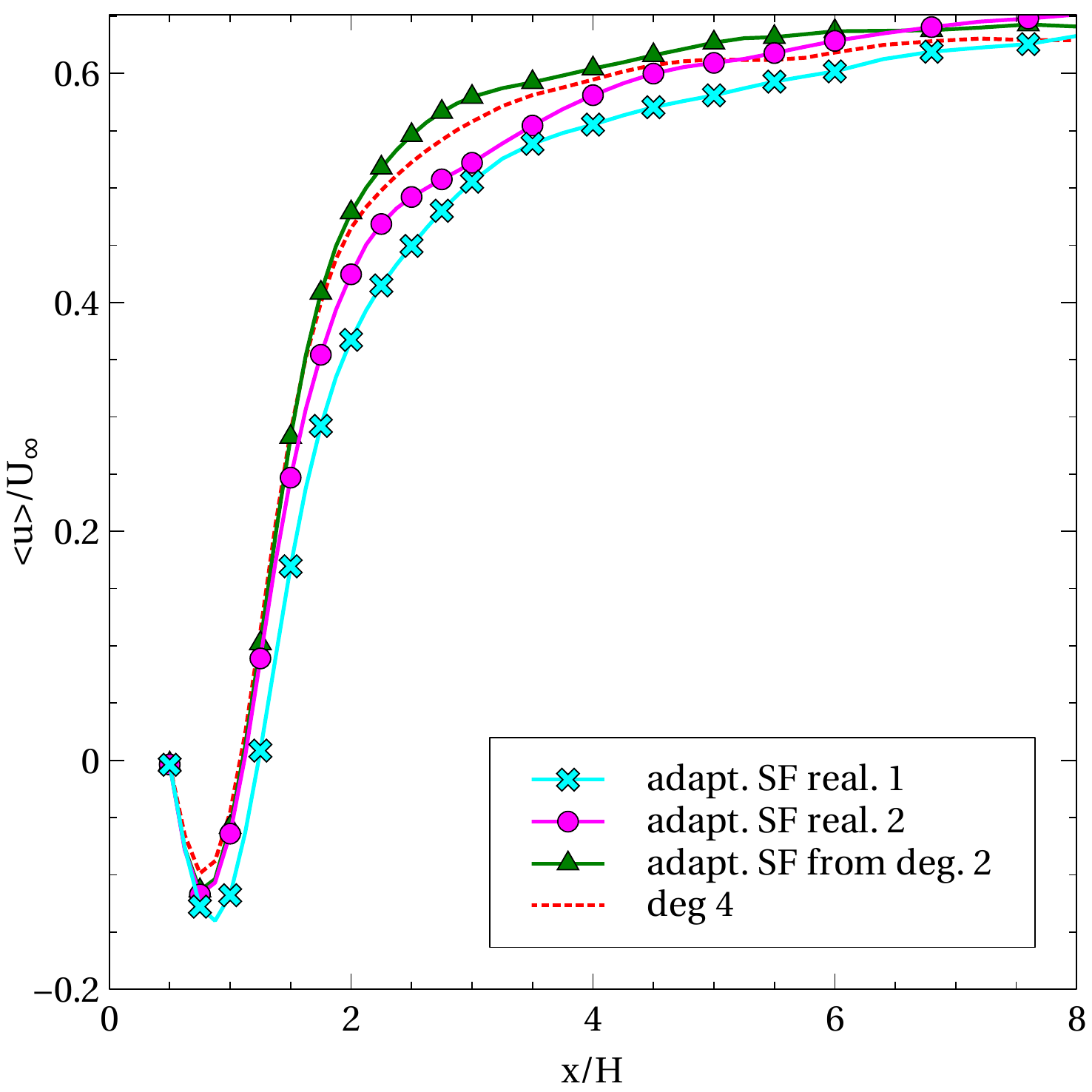}}   
\end{subfigure} \\
\begin{subfigure}[square root of total turbulent stresses, xx component]{\label{fig:adaptlow_h1_rmsu}
    \includegraphics[width=0.45\textwidth]{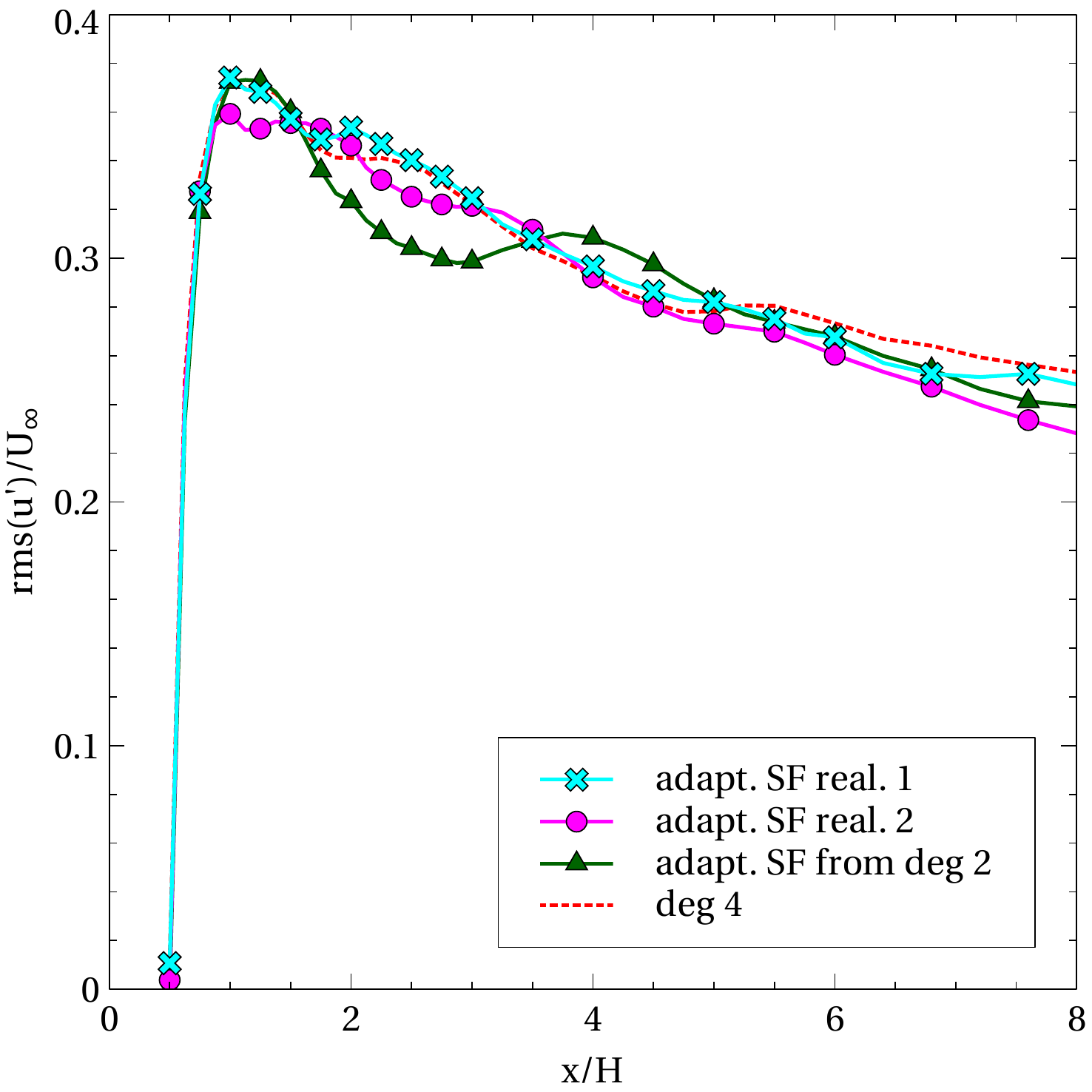}}    
\end{subfigure} 
\hfill
\begin{subfigure}[square root of total turbulent stresses, yy component]{\label{fig:adaptlow_h1_rmsv}
    \includegraphics[width=0.45\textwidth]{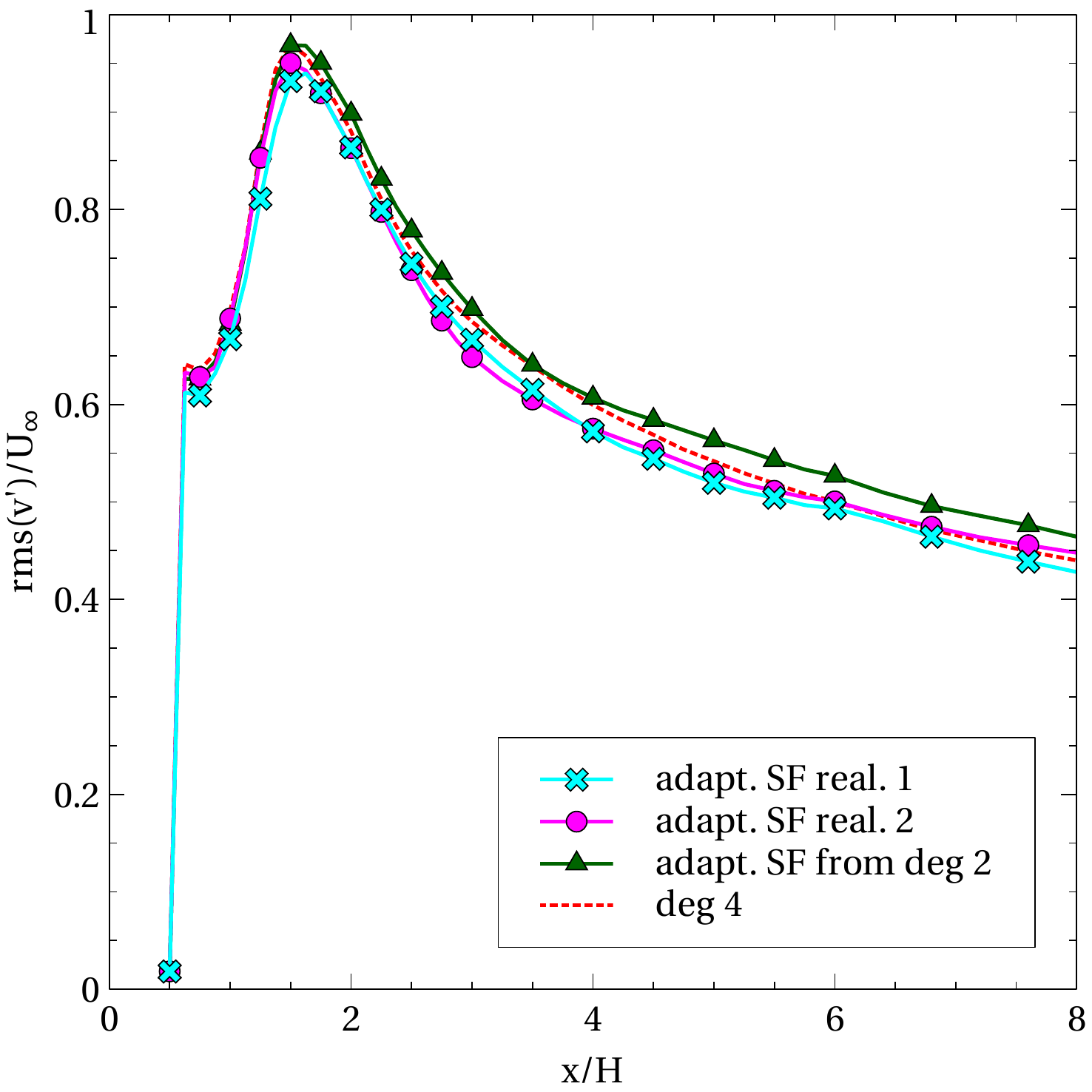}}    
\end{subfigure} 
\caption{Velocity statistics in the wake of the cylinder, along plot line Z (see fig. \ref{fig:cyl}), comparison of results based on indicators calculated from different data }
\label{fig:adaptlow_h1}
\end{figure*}

\begin{figure*}
\centering
\begin{subfigure}[Streamwise mean velocity]{\label{fig:adaptlow_v5_umean}
    \includegraphics[width=0.45\textwidth]{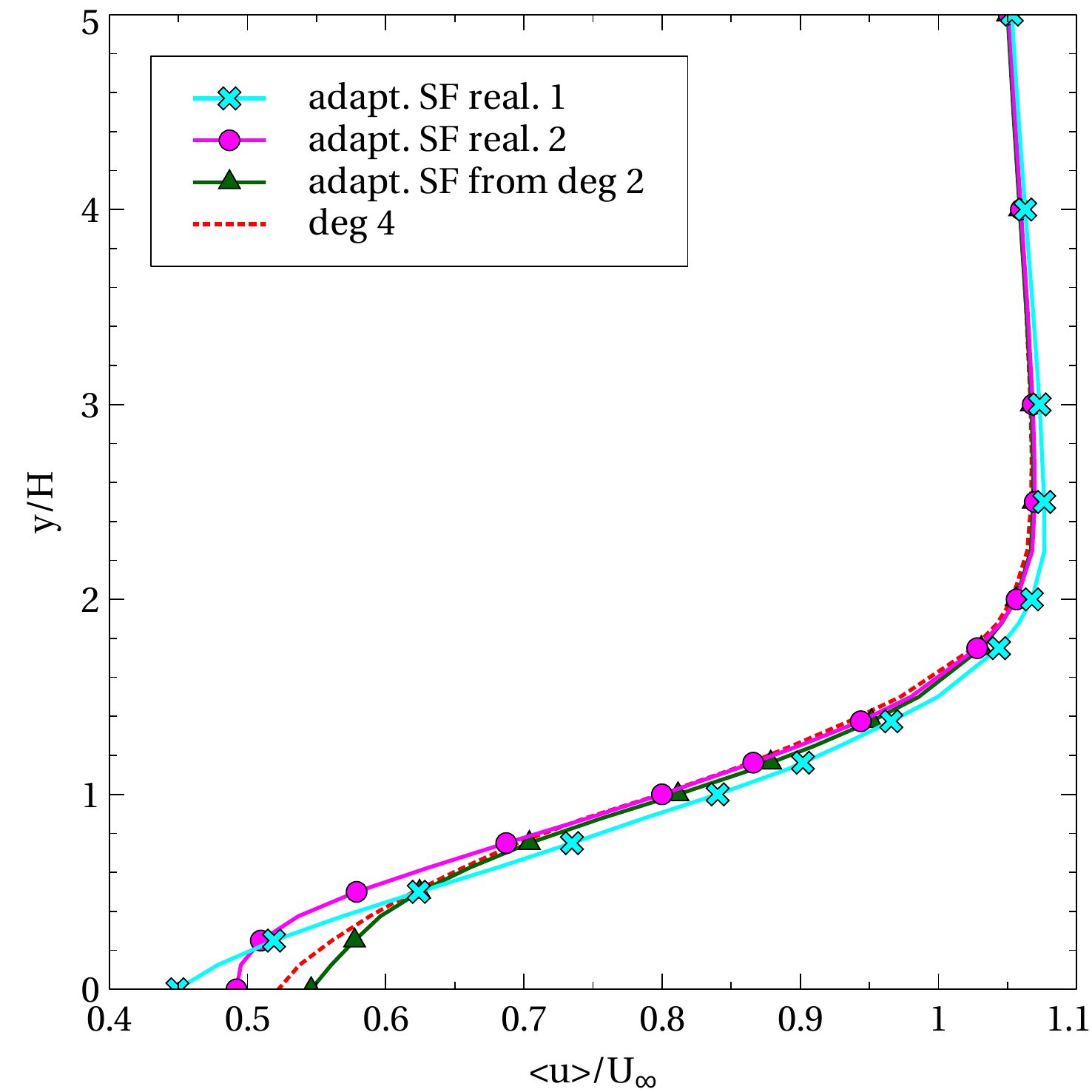}}    
\end{subfigure}\hfill 
\begin{subfigure}[total turbulent stresses, xy component]{\label{fig:adaptlow_v5_tauuv}
    \includegraphics[width=0.45\textwidth]{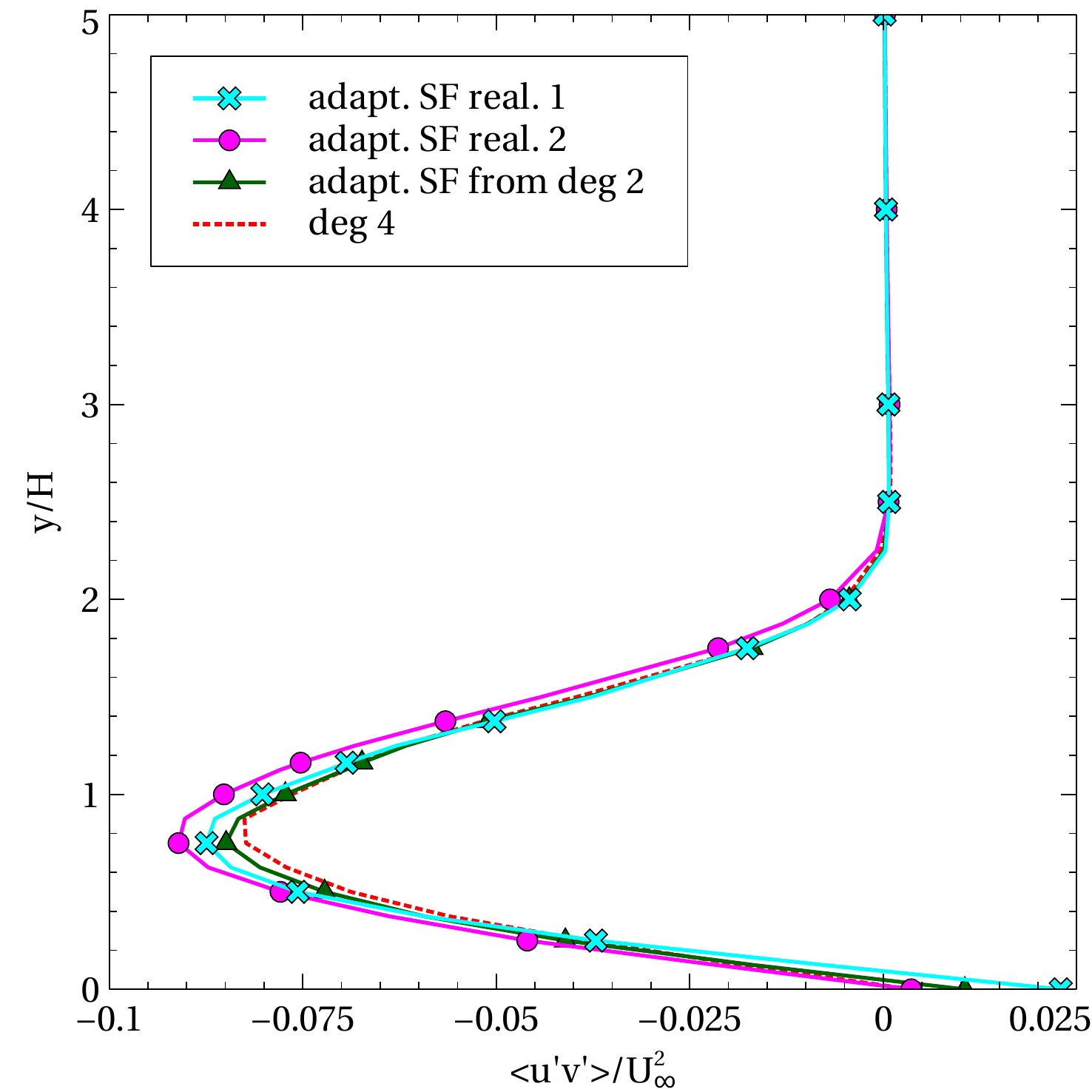}}    
\end{subfigure}\\
\begin{subfigure}[square root of total turbulent stresses, xx component]{\label{fig:adaptlow_v5_rmsu}
    \includegraphics[width=0.45\textwidth]{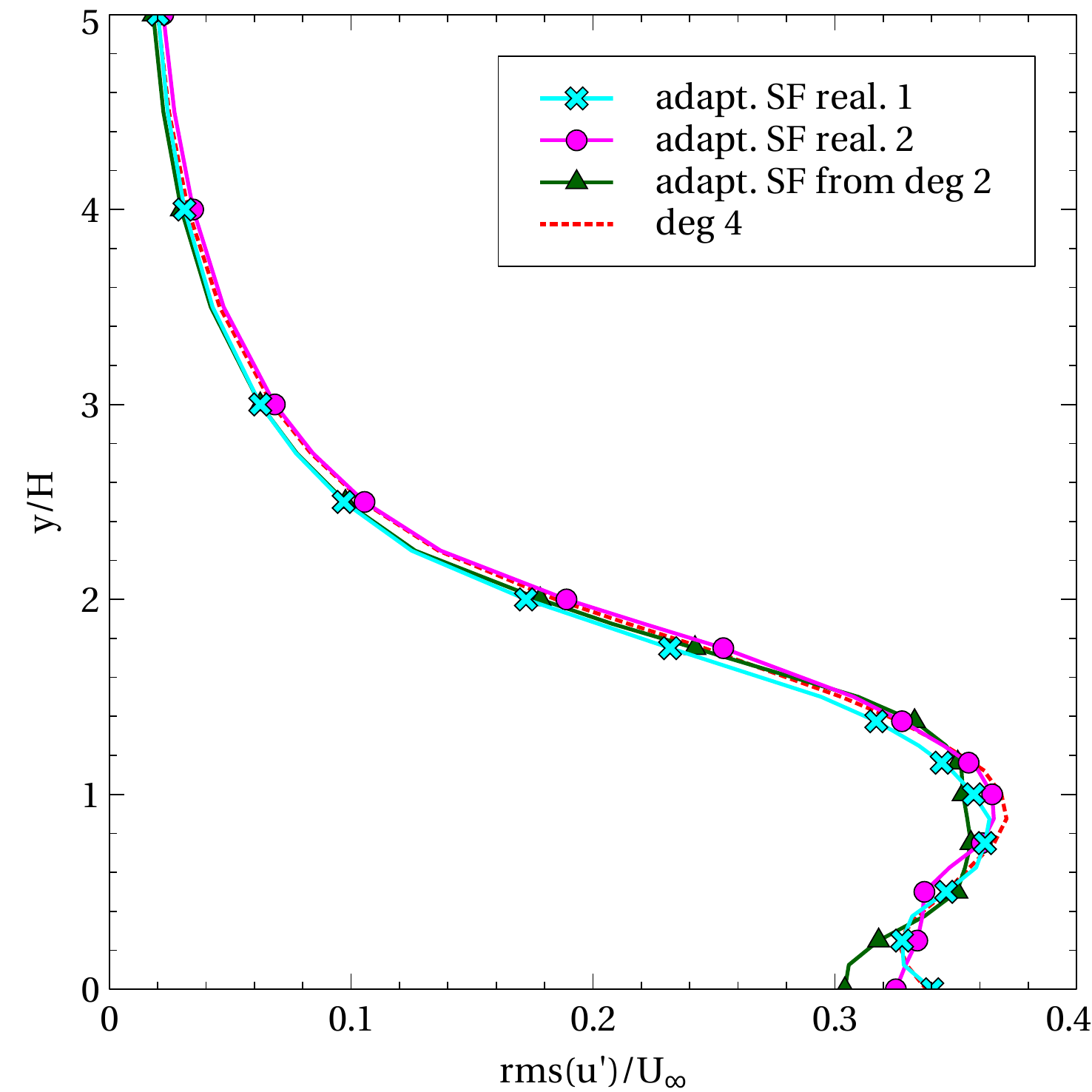}}    
\end{subfigure} 
\hfill
\begin{subfigure}[square root of total turbulent stresses, yy component]{\label{fig:adaptlow_v5_rmsv}
    \includegraphics[width=0.45\textwidth]{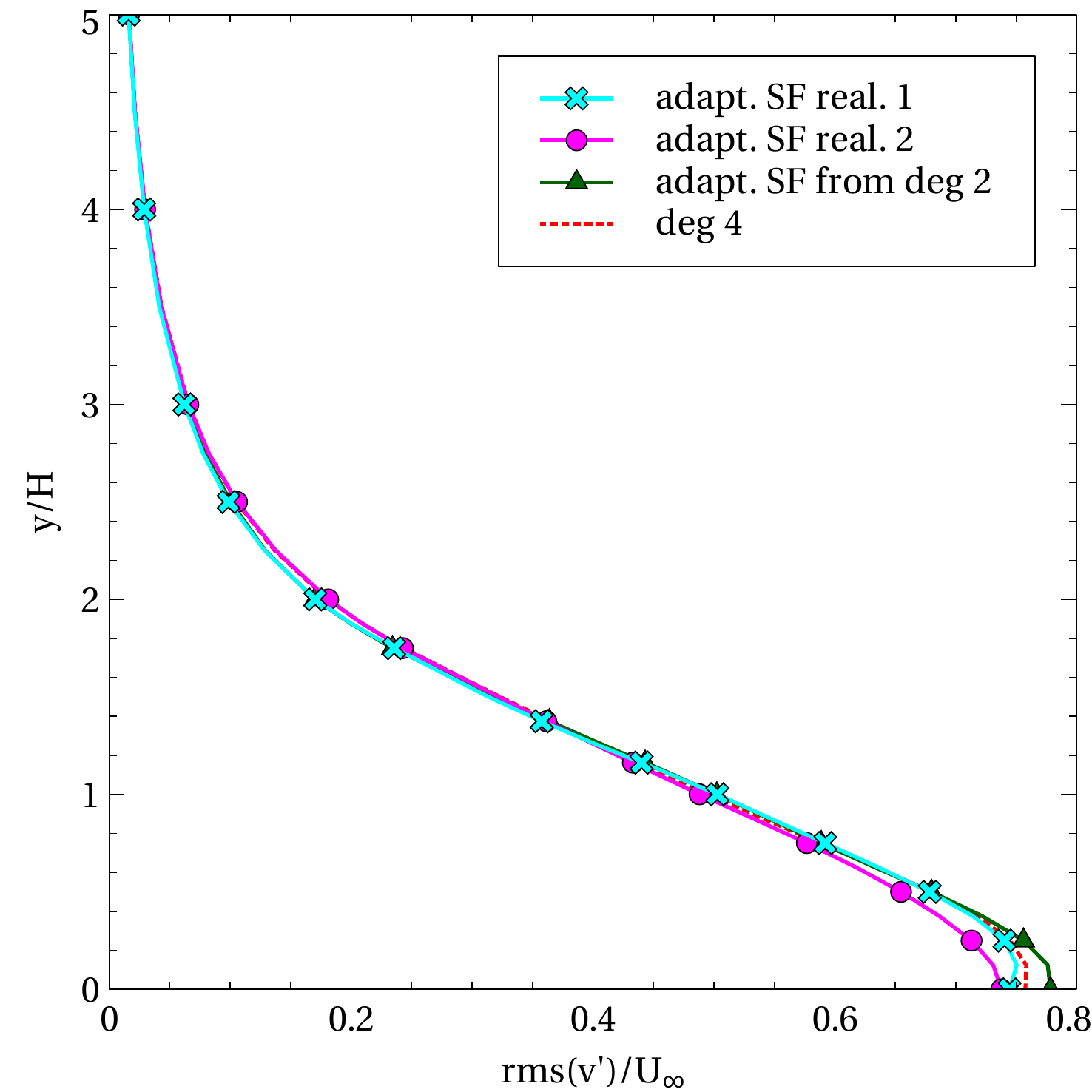}}    
\end{subfigure} 
\caption{Velocity statistics across the wake of the cylinder, along plot line Z (see fig. \ref{fig:cyl}), comparison of results based on indicators calculated from different data}
\label{fig:adaptlow_v5}
\end{figure*}

The global data are presented in table \ref{tab:adaptlow-global}, while velocity statistics profiles are shown in figures \ref{fig:adaptlow_h1} and \ref{fig:adaptlow_v5}.
Both in the global results and in the statistics profiles, the simulation performed with the indicator calculated from the two different realizations shows some variability in spite of the similar polynomial distribution. However, the differences between the results are clearly smaller with respect to those between simulations performed with lower constant polynomial degree and the reference simulation,
or those between results achieved with different adaptation indicators and the reference
simulation.   Therefore, the adaptive simulation does not appear to be especially sensitive to the preliminary run data. 
Even more interestingly, it can be seen that   the results obtained with the polynomial distribution calculated from low degree preliminary runs, fall within the same variability range as the results based on the degree 4 results.

\begin{table*}
\centering
\begin{tabular}{p{3cm}ccccccc}
\toprule 
                & St            & <Cd> & rms(Cd')    & rms(Cl)  & dofs & core h.    \\ \midrule
P4        & 0.1410      & 2.398           & 0.1930      & 1.374 & 833560 &  7561   \\
AP4R1       & 0.1410      & 2.350           & 0.1692      & 1.241 & 398270 &  2662    \\
AP4R2      & 0.1425      & 2.400           & 0.1486      & 1.380 & 397210 &  2560    \\
AP2        & 0.1483     & 2.390           & 0.1642      & 1.338 & 398105 & 2640  \\
\bottomrule
\end{tabular}
\caption{Global quantities from different  simulations with SF indicator for sensitivity analysis: P4) constant degree 4, AP4R1) adaptive simulation with SF criterion computed from degree 4 simulation,
 AP4R2) adaptive simulation with SF criterion computed from a different degree 4 simulation,
  AP2) adaptive simulation with SF criterion computed from degree 2 simulation.
}
\label{tab:adaptlow-global}
\end{table*}

This simulations show not only that the adaptation procedure is robust, but also that it is possible to obtain good results based on an indicator computed from a very coarse preliminary simulation, which itself is inexpensive and produces unsatisfactory results as shown in figures \ref{fig:ref_h1} and \ref{fig:ref_v5}. The coarse simulation contains however enough information to allow an effective adaptation and accurate results from the following adaptive simulation. The total core hours used to obtain the results from degree 2, starting from zero, is 3682, comprising the time needed to achieve a statistical stationary state, accumulate statistics at constant degree 2, a buffer time in the adapted configuration to allow the transition at the new degree distribution and the statistics accumulation time in the adapted configuration. 
This figure is around just $38\%$ of the 9778 core hours needed to obtain the full degree 4 polynomial base results starting from zero, and shows that the static adaptation procedure is not only effective in terms of quality of the results, but allows for a substantial computational effort reduction also when approaching a new case without refined reference results, which is the most realistic situation.

\section{Conclusions and future perspectives}
\label{sec:conclu} \indent
We have introduced and tested a novel approach for
 adapting the local polynomial degree in the DG discretization of a compressible LES model. 
The proposed indicator  tries to measure the local turbulence intensity in the flow
by computing an approximation of the structure function, rather than 
computing a local  error estimate.  The novel indicator performance has been compared
to that of more conventional adaptation criteria using the flow around a bluff body as benchmark.
  Static adaptivity based on preliminary model runs has been performed. 
Both indicators were capable of highlighting domain areas of major turbulent activity, but the indicator based on the evaluation of the structure function proved itself more capable to lead to accurate results than the one based on the relative weight of the modal solution. 
Results obtained with the  novel indicator led to accurate results, comparable to those obtained
with constant maximum polynomial degree, with a reduction of the computational effort 
of approximately 60\%. A sensitivity analysis has also been carried out, showing that
an accurate adaptive solution can be achieved using a relatively coarse resolution in the preliminary runs, thus outlining a practical procedure to obtain efficient adaptive results with minimal additional effort.

The adaptation procedure in the LES context has been shown to be effective, in spite of the complexities and peculiarities illustrated in section \ref{sec:intro}. The use of physically based indicators produced satisfactory results and proved to be a good candidate to overcome the difficulties related
 to adaptive LES. 
Given this encouraging results,  on one hand we plan to keep investigating also other indicators capable to easily assess the local flow conditions from the LES viewpoint. On the other hand, 
 we are currently developing a dynamic adaptation procedure based on the same approach proposed in this paper. In this way we hope to extend the benefits of adaptation to statistically non stationary phenomena, such as vortex interaction, moving boundaries and general transient phenomena.

 \section*{Acknowledgements} 
We would like to thank Florian Hindenlang for the useful discussions on adaptivity algorithms and dynamic adaptivity.

We acknowledge that the results of this research have been achieved using the 
computational resources made available at the CINECA supercomputing center (Italy) by the high 
performance computing project ISCRA-C HP10CHV1QD.

\begin{appendix}
\section{Structure function form in homogeneous isotropic turbulence}
\label{app:anisotropy-calc}
We report here the calculations necessary to obtain an analytical expression of the isotropic form of the structure function from a calculated structure function, and then to evaluate the difference between the isotropic form and the calculated structure function.
Given the structure function 
$$
	D_{ij}(\rvect, \xvect, t) = \mean{\left[U_i(\xvect+\rvect,t)-U_i(\xvect,t)\right]\left[U_j(\xvect+\rvect,t)-U_j(\xvect,t)\right]}
	$$
	
and its corresponding expression in the isotropic case
\begin{equation}
\label{eq:iso_strucfun-app}
	D_{ij}^{iso}(\rvect, t) = D_{NN}(r,t)\delta_{ij} +\left(D_{LL}(r,t) - D_{NN}(r,t)\right)\frac{r_ir_j}{r^2},
\end{equation}
it is possible to define a quadratic error 
\begin{equation}
\label{eq:qerr_def-app}
Q = \left[D_{ij}^{iso}-D_{ij}\right]^2 = \left[D_{ij}^{iso}-D_{ij}\right] \left[D_{ij}^{iso}-D_{ij}\right]
\end{equation}
which, thanks to the definition \ref{eq:iso_strucfun-app} becomes:
\begin{equation}
\label{eq:qerr_form-app}
Q = \left[D_{NN}(r,t)\delta_{ij} +\left(D_{LL}(r,t) - D_{NN}(r,t)\right)\frac{r_ir_j}{r^2}-D_{ij}\right]^2.
\end{equation}
Looking for the values of $D_{LL}$ and $D_{NN}$ which minimize the error, the following system of equations must be solved:
$$ \frac{\partial Q}{\partial D_{NN}} = 0, \ \ \  \frac{\partial Q}{\partial D_{LL}} = 0, $$
which becomes
$$\left[D_{NN}(r,t)\delta_{ij} +\left(D_{LL}(r,t) - D_{NN}(r,t)\right)\frac{r_ir_j}{r^2}-D_{ij}\right] \left(\delta_{ij} - \frac{r_ir_j}{r^2}\right)= 0 $$
$$
\left[D_{NN}(r,t)\delta_{ij} +\left(D_{LL}(r,t) - D_{NN}(r,t)\right)\frac{r_ir_j}{r^2}-D_{ij}\right] \left(\frac{r_ir_j}{r^2}\right)= 0. $$
Noting that 
$ \delta_{ij}\delta_{ij} = 3, $
$\delta_{ij} A_{ij} = \tr{A_{ij}} $ and
$$\tr{\frac{r_ir_j}{r^2}} = \frac{r^2}{r^2} = 1, $$
the final formulation becomes
$$D_{NN} = \frac{D_{ij}\frac{r_ir_j}{r^2} - \tr{D_{ij}}\left( \frac{r_ir_j}{r^2}\right)^2}{1-3\left( \frac{r_ir_j}{r^2}\right)^2}$$
$$D_{LL} = D_{NN}\left( \tr{D_{ij}} -2 \right). $$
Finally, once   the two coefficients $D_{NN},$ $D_{LL} $  have been computed, it is possible to evaluate the quantity $Q $ using equation \ref{eq:qerr_form-app}.

\end{appendix}

\bibliographystyle{plain}
\bibliography{dg_les}
\end{document}